\documentclass[acmsmall]{acmart}

\usepackage{amsmath,amsfonts}
\usepackage{siunitx}
\usepackage{algorithm, algorithmicx, algpseudocode}
\usepackage{graphicx}
\usepackage{textcomp}
\usepackage{multirow}
\usepackage{xcolor}
\usepackage{makecell}
\usepackage{tcolorbox}
\usepackage{tabularx, threeparttable}
\usepackage{soul}
\usepackage{booktabs}
\usepackage{hyperref}
\usepackage[caption=false]{subfig}
\usepackage{enumitem}
\usepackage{arydshln}
\usepackage{bm}
\usepackage{xcolor}
\usepackage[normalem]{ulem}

\newif\ifshowcomments
\showcommentsfalse 

\definecolor{lwbcolor}{RGB}{247,242,34}  
\definecolor{lwbcolorbg}{RGB}{24,117,210}  

\newcommand{\modify}[2]{#2}

\newcommand{\add}[1]{#1}

\definecolor{myblue}{RGB}{0,112,192}
\definecolor{myred}{RGB}{192,0,0}

\newcommand{\modifycolor}[1]{#1}

\makeatletter
\renewcommand{\Statex}{\noindent\hskip-\ALG@thistlm}

\makeatother
\AtBeginDocument{%
  }

\setcopyright{cc}
\setcctype{by}
\acmDOI{10.1145/3808185}
\acmYear{2026}
\acmJournal{PACMSE}
\acmVolume{3}
\acmNumber{FSE}
\acmArticle{FSE178}
\acmMonth{7}
\acmSubmissionID{fse26mainb-p1647-p}
\received{2026-02-24}
\received[accepted]{2026-03-24}

\begin{document}

\title{Failure-Based Testing for Deep Reinforcement Learning Agents}

\author{Weibin Lin}
\orcid{0009-0001-7105-9312}
\affiliation{%
  \institution{Beihang University}
  \city{Beijing}
  \country{China}
}
\email{linwb@buaa.edu.cn}

\author{Jiangtao Meng}
\orcid{0009-0007-5384-998X}
\affiliation{%
  \institution{Beihang University}
  \city{Beijing}
  \country{China}
}
\email{mengjiangtao@buaa.edu.cn}

\author{Zheng Zheng}
\orcid{0000-0001-7922-9067}
\affiliation{%
  \institution{Beihang University}
  \city{Beijing}
  \country{China}
}
\email{Zhengz@buaa.edu.cn}
\authornote{Corresponding author}

\begin{abstract}
Deep Reinforcement Learning (DRL) agents have been widely adopted across diverse domains to address challenging decision-making problems, such as autonomous driving and robotic control. Given that many of these applications are safety- and security-critical, rigorous testing of DRL agents is indispensable. Existing testing methods are typically guided by reward signals to detect failures. However, for well-trained agents, whose performance approaches optimal levels in standard operating conditions, reward signals remain generally high, making current methods ineffective at uncovering critical failures.

To address these challenges, we propose a novel failure-based method that leverages task-induced failure insights to enhance failure detection capability while reducing the number of tests required. Since DRL agents are inherently designed with human-defined tasks, they provide valuable cues about task difficulty. Intuitively, a DRL agent is more likely to fail when confronted with a more difficult task; therefore, PRT prioritizes these tasks. Building on this foundation, we propose Prior Random Testing, a black-box failure-based testing method that enables targeted prioritization while preserving the diversity of generated test cases. Guided by task-induced failure insights, PRT prioritizes failure-prone regions of the input domain, thereby facilitating efficient failure detection.

PRT is evaluated on \modify{three}{four} widely used benchmarks and compared with different state-of-the-art methods including fuzzing, search-based and generative-based methods. PRT ranks among the top performers in terms of both the cost of finding the first failure and the diversity of test cases. \add{Notably, compared to random testing, PRT achieves better diversity and reduces the testing cost by over 50\%.}
\end{abstract}



\maketitle

\section{Introduction}

Artificial intelligence (AI) has shown impressive performance in solving sequential decision-making tasks using deep reinforcement learning (DRL) techniques. DRL agents are widely employed in complex tasks such as autonomous driving \cite{ad}, robot control \cite{rc}, traffic control \cite{traffic}, and gaming \cite{game1, game2}. Despite these advances, the reliability of DRL agents remains uncertain \cite{a17070314}. Wrong decisions made by agents could lead to catastrophic consequences, particularly in safety-critical domains \cite{10430398}. This growing reliance on DRL agents underscores the urgent need to rigorously test these agents to ensure their quality.

Real-world incidents highlight this necessity. For instance, a Tesla autonomous taxi was recently involved in its first recorded accident, colliding with a parked car \cite{tesla}. Although commercial products are generally released by well-trained individuals, even rare failures can be disastrous. Therefore, efficiently testing almost flawless DRL agents is particularly important.

Existing methods that can be applied to test different types of DRL agents are typically guided by reward signals \cite{mdpfuzz,curefuzz,qd}. Reward-based guidance exploits the property that low rewards are often correlated with failures. However, when agents demonstrate well-trained performance, rewards are consistently high, offering little useful signal. For example, in the Cart Pole environment \cite{cartpole}, all successful test cases output exactly the same reward, providing no distinction for guiding testing. 

When reward signals provide limited guidance, failure-prone regions of the input domain can often be identified more directly. Since DRL agents are inherently task-driven and defined by human-specified oracles, it is possible to distinguish between easier and more difficult tasks, where the latter are more likely to induce failures. We define this intuition as task-induced failure insights, referring to the identification of input domain locations corresponding to the most difficult tasks within a given environment.
A natural concern, however, is whether testing only the most difficult tasks suffices. Eniser et al. \cite{eniser2022metamorphic} show that an agent may succeed in a difficult task yet fail in an easier one. \modify{This observation underscores the need to balance prioritization with broad exploration: while difficult tasks are more failure-prone, uniform coverage of the input domain remains necessary.}{This observation underscores broad exploration: when the location of failure-prone regions is unknown a priori, evenly distributed test cases over the input space become an effective surrogate strategy for failure discovery, serving as a principled fallback when testing the most difficult tasks fails to expose failures.}

This naturally connects to the idea of failure-based testing, first introduced by Chen et al. \cite{art2}. Failure-based methods select test cases based on prior knowledge of failure patterns, including geometric shapes, sizes, or locations of failure regions \cite{art2}. Adaptive Random Testing (ART) \cite{art} is a representative example that \modify{exploits failure size information. Unfortunately, for DRL agents, the size of failure regions is typically unknown. By contrast, location information can often be approximated from task-induced insights, although such information may be imprecise.}{assumes that failure-causing test cases form contiguous regions, while passing test cases also exhibit clustering behavior. Therefore, given a set of previously executed test cases that have not exposed failures, selecting new test cases farther from these existing ones increases the likelihood of discovering failures \cite{art2}. While preserving this sparsity-driven exploration strategy, we further leverage failure location information to guide failure-based testing.} 

Building on this, we propose Priority Random Testing (PRT), a black-box failure-based test case generation method that enables prioritization with uniformity. Before testing, PRT needs to design the hyperparameters in \modify{two}{three} steps:
\modify{
  (1) Estimate failure-prone locations by leveraging task-induced failure insights for each dimension of the input domain;
  (2) Adjust for uncertainty by assigning a confidence-level hyperparameter to the estimated locations.
}{
  (1) \textbf{Failure-prone Region:} For every dimension of the input domain, identify the failure-prone value for prior testing, here corresponding to the most difficult tasks. 
  (2) \textbf{Mapping $\mathcal{F}$:} Design a mapping $\mathcal{F}(\cdot)$ for every dimension, mapping the boundary to the failure-prone value. 
  (3) \textbf{Confidence $\lambda$:} Define the certainty level for that failure-prone value. A larger value of $\lambda$ indicates a higher prior confidence in the failure-prone designation.
}

PRT analytically generates new test cases within the sparsest regions of existing test cases to ensure uniformity, \modify{with}{naturally resulting in} a default priority on boundary exploration. \add{For example, in one-dimensional input domain, regardless of the location of the first point, the farthest point from it (i.e., the sparsest location) necessarily lies at the boundary.} 

\add{We briefly introduce the workflow of PRT as follows. Assume that we already have a test case set $\mathcal{C}=\{\varepsilon_1, \varepsilon_2, \cdots, \varepsilon_n\}$.} When generating a new test case $\varepsilon_{n+1}$, PRT decomposes the problem across dimensions and generates values dimension by dimension. \modify{It first identifies the sparsest dimension and produces the corresponding value, a process we term \textit{Dimension Reduction}, since only the remaining dimensions need to be considered afterward.}{It first computes the largest interval $E_i$ of each dimension $i$ in the set $\mathcal{C}$. It then selects the sparsest dimension $k=\arg\max_i(E_i)$ and generates the corresponding solution $\varepsilon_{n+1}^k$ (the optimal solution for the $k$-dimensional largest interval $E_k$). We term this process \textit{Dimension Reduction}.} \modify{Next, PRT filters existing test cases whose distances exceed a threshold $\tau$ (a hyperparameter) along the dimensions already generated for $\varepsilon$.}{Next, PRT filters existing test cases to a set $\mathcal{C}'=\{x\in \mathcal{C} \mid x^k \text{ close to } \varepsilon_{n+1}^k\}$.} This step, is referred to as \textit{Local Recombination}\modify{, further strengthens boundary-prior exploration}{}. By alternating \textit{Dimension Reduction} and \textit{Local Recombination}, a new test case $\varepsilon_{n+1}$ is obtained.
\modify{In addition, we incorporate a linear mapping to shift boundary regions into other areas of the domain, enabling the customization of prior regions according to different failure insights. We realize PRT with a low computational complexity, approximating $O(MN^2)$ for generating $N$ test cases in an $M$-dimensional domain.}{In addition, we incorporate a mapping $\mathcal{F}$ to shift boundary regions into other areas of the domain. Though PRT prioritizes testing the boundary by default, mapping $\mathcal{F}$ enables customized prioritization of other regions according to different failure insights. We implement PRT with a time complexity of approximately $O(MN^2)$ for generating $N$ test cases in an $M$-dimensional domain.}

In summary, this paper makes the following contributions:

\begin{itemize}[leftmargin=1em]
  \item We propose a failure-based method, PRT, with a time complexity $O(MN^2)$, capable of leveraging the task-induced failure insights on DRL agents to efficiently find failures. 
  \item We evaluate PRT against four state-of-the-art (SOTA) methods---including fuzzing, search-based, and generative-based methods---on three representative testing subjects with distinct reward structures. PRT consistently ranks among the top performers in terms of both effectiveness and efficiency for detecting the first failure while achieving the best diversity of test cases.
  \item We further empirically investigate the failure patterns of different DRL agents to evaluate the effectiveness and generalizability of PRT. Our findings reveal two main types of patterns: block-shaped and point-shaped. PRT addresses these through task-induced failure insights and uniform distribution, respectively. 
\end{itemize}

\section{Background}

\subsection{Preliminaries}
\label{Preliminaries}

\textbf{Markov Decision Process (MDP).} MDP is a mathematical framework used for modeling decision-making, which is composed of a tuple $<S,A,\mathcal{T},\mathcal{R},\pi>$. $S$ is a set of all possible states during MDP. $A$ is a set of actions available to the decision-making agent. $\mathcal{T}(s_{t+1} |s_t,a_t)\rightarrow[0,1]$ defines the probability from the present state $s_t$ to the next state for every $s_{t+1}\in S$ with action $a_t$. The reward function $\mathcal{R}(s_t, a_t)\rightarrow\mathbb{R}$ is used to evaluate the quality of an action $a_t$ under a specified state $s_t$. Specifically, an episode reward means the cumulative reward from the starting state $s_0$ to the end of the episode. $\pi(a_t|s_t)\rightarrow[0,1]$ gives the probability of taking action $a_t$ under the state $s_t$, representing the behavior of an agent. 

\textbf{Deep Reinforcement Learning (DRL).} Based on the mathematical MDP model, DRL gives a solution of $\pi(a_t|s_t)$ \cite{reinforcementlearning}. Unlike deep learning, which directly labels an action as right or wrong, RL only scores an action with a reward value and lets the agents learn how to obtain the cumulative reward in a task as high as possible. The environment in RL refers to the tuple $<S,A,\mathcal{T},\mathcal{R}>$, and the agent refers to the policy $\pi$. The environment is responsible for giving the observed state and the corresponding reward to the agent, and the agent gives an action to the environment for the next state transition. As the deep neural network (DNN) demonstrates excellent performance in many areas, DRL employs DNN as the policy function $\pi(a_t|s_t)$. Popular DRL techniques include DQN \cite{dqn}, PPO \cite{ppo}, TQC \cite{tqc}, and so on.  

\textbf{Test Cases for DRL}. Though our testing subject, the DRL agent, takes states as inputs, our test cases are defined as the initial configurations of the environment. Once the agent takes the first action, we will not change anything, so as to ensure the authenticity of the scenario. 

\textbf{Input Domain.} The input domain is defined as the set of all feasible test cases. It is emphasized that the test case is not the input $s$ for testing subjects.

\textbf{Test Oracle.} We give a setting of the environment as the test case at the beginning. Then the agent interacts with the environment. If the state $s_t$ satisfies some conditions (depended on the environment), the test case fails or passes at the $t$ step.

\textbf{Failure Pattern.} We refer to the failure pattern as the characteristics of the failure test case distribution in the input domain, especially the geometric ones after visualization. 

\subsection{Task-Induced Failure Insights on DRL Agents}
\label{expertknowledge}
A concern is whether task-induced failure insights can generalize across different DRL agents. We demonstrate that such insights are broadly applicable because they originate from properties shared by all DRL agents. Specifically, every DRL agent is designed to solve a man-made task, and these tasks inherently contain cues about difficulty \cite{10.5555/3312046}. \modify{Furthermore, reward serves as a quantitative manifestation of the task, and training in DRL necessarily relies on reward signals. The reward design itself provides valuable information that can be leveraged when testing. Together, the task structure and its associated reward function form a universal basis for deriving failure insights, regardless of the specific agent.

}{}The core intuition is that agents are more likely to fail in more difficult tasks, with difficulty intuited directly from the task definition. For example, in games, difficulty can be increased by adding more obstacles or enemies, while in control problems, test cases can be designed farther from the stable state. \modify{Reward signals complement these cues by indicating whether a state is encouraged or discouraged, thereby helping to identify regions with higher failure probability.}{}

Amal et al. make a comprehensive review on testing DRL agents and summarizes the testing environments \cite{overview}. We select the environments tested by more than two papers and annotate their tasks along with the failure insights on the failure patterns in Table.~\ref{tab:expertknowledge}, so as to show that we can easily locate the failure-prone regions of the input domain. We also attach the cited numbers counted by Amal et al. \cite{overview}. 

\vspace{-0.2cm}
\begin{table}[h!]
  \begin{center}
    \caption{Task-induced insights about failures in different environments \cite{overview}}
    \label{tab:expertknowledge}
    \begin{threeparttable}
      \begin{tabularx}{1.0\textwidth}{ >{\centering\arraybackslash}p{2.2cm} X p{3.3cm} >{\centering\arraybackslash}p{0.8cm} }
        \hline
        Environment & Task & Insights\tnote{1} & Papers \\
        \hline
        Frozen Lake & Walk from start to goal without falling into any holes & More holes & 3 \\
        Mountain Car\tnote{2} & Accelerate a car from the bottom of a sinusoidal valley to the top of the right hill & Starting from the middle with no velocity & 8 \\
        Cart Pole\tnote{2} & Balance a pole attached by an un-actuated joint to a cart & Higher initial velocity & 12 \\
        Lunar Lander\tnote{2} & Safely land on the middle pad from the sky & Higher initial velocity & 7 \\
        Super Mario & Arrive at the destination without collision with the enemies or falling into the pits & More enemies and pits & 3 \\
        Coop Navi (Simple Spread) & Several agents need to cover the same number of landmarks while avoiding collisions & Agents are closer to each other & 3 \\
        Bipedal Walker & Balance a 4-joint robot to step over obstacles & Larger size of obstacles & 4 \\
        CARLA & Control a car to arrive at the destination without collision & More vehicles on roads & 4 \\
        Taxi & Navigate a car to pick up a passenger and drop him off at the destination & The locations are far from each other & 3 \\
        \hline
      \end{tabularx}
      \small \item [1] Insights indicate  how to design a test case with a higher failure probability.
      \small \item [2] For these three environments, we introduce more details in Sec. \ref{testingsubjects}.
    \end{threeparttable}
  \end{center}
\end{table}

\vspace{-0.6cm}
\section{Approach}
PRT consists of two cooperative mechanisms: dimension reduction and local recombination. \add{Dimensionality reduction provides a global view that identifies sparse regions among the candidate set, while local recombination examines which region is still dense and offer that as the new candidate set.} \modify{Every time generating a new test case $\varepsilon$, PRT splits the test case dimension by dimension (dimension reduction). Each time PRT selects the sparsest dimension (assumed to be the $k$-th dimension) from a candidate set $\mathcal{C}$ to generate the corresponding solution $value$. After generating a dimension's solution $\varepsilon^k$, PRT reconstructs the candidate set $\mathcal{C}$ by only considering the test cases in $\mathcal{C}$ close to $\varepsilon^k$ in the $k$-th dimension (local recombination).}{Each time PRT generates a new test case $\varepsilon$, it performs dimension reduction by splitting the test case space along individual dimensions: $\varepsilon=(\varepsilon^1,\varepsilon^2,\cdots,\varepsilon^m)$. Each time, PRT selects the sparsest dimension (assumed to be the $k_1$-th dimension) from the candidate set $\mathcal{C}$ to generate the corresponding optimal solution $value$ so that $\varepsilon^{k_1}=value$. Then PRT reconstructs a candidate set $\mathcal{C}_1\subset\mathcal{C}$ by only including the test cases in $\mathcal{C}$ close to $\varepsilon^{k_1}$ in the $k_1$-th dimension (local recombination). Next round we will have $\varepsilon^{k_2}$ assigned to the $value$ corresponding to the sparsest $k_2$-th dimension, and a smaller candidate set $\mathcal{C}_2\subset\mathcal{C}_1$, and so on up to $k_m$, where $\{k_1,k_2,\cdots,k_m\}=\{1,2,\cdots,m\}$. At last we can obtain a new test case $\varepsilon$.}

\vspace{-0.45cm}
\begin{equation}
  FP(InD)=\{fp^1,fp^2,\cdots,fp^m\}
  \label{eq:fp}
\end{equation}

By default, PRT prioritizes testing the boundary. \modify{We need to design a mapping to shift the locations to failure-prone regions according to the task-induced failure insights.}{However, the region we want to prioritize (corresponding to the most difficult tasks) does not always lie in the boundary. At the beginning of applying PRT, we need to identify the failure-prone ($FP$) region of the input domain ($InD$) as Eq.~\ref{eq:fp}.}

\noindent\modify{More specifically, for every element of test cases (i.e., every dimension of the input domain), we need to specify the value corresponding to the most difficult task.}{More specifically, for every dimension of the input domain $InD^i$, we need to specify the failure-prone value $fp^i$.} By default, the value is set as the boundary, both the left side and the right side. We can change PRT's exploring strategy by a well-designed mapping $\mathcal{F}$, shifting the prior testing region. We regard $\mathcal{F}$ to be a hyperparameter as it is manually defined before testing. After setting the other hyperparameters, we can begin testing with PRT. 

\vspace{-0.2cm}
\subsection{Approach Overview}

Alg.~\ref{algo:prt} shows the workflow of PRT. It takes the number of test cases $N$ and \add{test case} dimension $M$ as inputs along with hyperparameters $\mathcal{F}, \lambda\in [1,+\infty)$, and gives the generated test case set $\mathbb{T}$ as the output. Line 2 randomly generates the first test case $\varepsilon$ as the initial reference. Line 3 constructs the output test case set $\mathbb{T}$. \modify{Lines 4--14 generate a test case $\varepsilon$ at each round. Lines 7--12 generate a dimensional value for $\varepsilon$ according to the set $\mathcal{C}$ at each round. Each time line 8 picks the sparsest dimension $k$ (with the largest interval) and generates the corresponding value, which is assigned to $\varepsilon^k$ at line 9. Line 10 uses such information to reconstruct a subset from $\mathcal{C}$. As we have generated $k$-dimensional data, line 11 reduces the dimension. Until all dimensional data have been generated, we can obtain a new test case $\varepsilon$.}{The outer \textbf{while} loop, lines 4--17, generates a new test case $\varepsilon$ at each round, where line 16 adds it to the output test case set $\mathbb{T}$. Line 5 creates a candidate set, initialized as the existing test case set $\mathbb{T}$. Line 6 creates an empty container for the generating test case $\varepsilon$ and Line 7 assigns an adaptive hyperparameter for the inner \textbf{while} loop. Line 8 creates an index set for indicating which dimensional value of $\varepsilon$ has not been generated. The inner \textbf{while} loop, lines 9--15, generates a dimensional value for $\varepsilon$ at each round. Each round line 10 picks the sparsest dimension $k$ (i.e., with the largest interval) from the candidate set $\mathcal{C}$ and generates the corresponding value resulting from the largest interval, which is assigned to ${\varLambda[k]}$-dimensional of $\varepsilon$ (i.e., $\varepsilon^{\varLambda[k]}$) at line 11. Line 12 uses such information to select test cases similar to $\varepsilon^{\varLambda[k]}$, where the new candidate set $\mathcal{C}$ is a subset of the old one. As we have generated $k$-dimensional data, line 13 reduces that dimension of $\mathcal{C}$ and the index set $\varLambda$ removes that dimension (e.g., remove the second dimension of $\{(1,2,3),(4,5,6)\}\rightarrow\{(1,3),(4,6)\}$).} Until all dimensional data have been generated, we can obtain a new test case $\varepsilon$. Lines 18--20 make a mapping $\mathcal{F}$ on every test case in $\mathbb{T}$.

\vspace{-0.2cm}
\begin{algorithm}[h!]
\modifycolor{
  \caption{Prior Random Testing}
  \label{algo:prt}
  \setlength{\baselineskip}{1pt}
  \begin{algorithmic}[1]
    \Require expected number of test cases $N$, test case dimension $M$, mapping $\mathcal{F}$, confidence $\lambda$
    \Ensure test case set $\mathbb{T}$
    \Function{PRT}{$N,M,\mathcal{F},\lambda$}
      \State $\varepsilon_{random}$$\leftarrow$ randomly generate an M-dimensional test case 
      \State $\mathbb{T}\leftarrow\{\varepsilon_{random}\}$
      \While{$|\mathbb{T}| < N$} \Comment{$|\mathbb{T}|$: the number of $\mathbb{T}$}
        \State $\mathcal{C}\leftarrow copy(\mathbb{T})$
        \State $\varepsilon\leftarrow$ an M-dimensional empty array \Comment{the next test case}
        \State $\tau\leftarrow (|\mathbb{T}|+1)^{-1/M}$ \Comment{an adpative hyperparameter for L\textsc{ocal}R\textsc{ecombination}}
        \State $\varLambda \leftarrow \{1,2,\cdots,M\}$ \Comment{a dimensional index set}
        \While{$\varLambda$ is not empty}
          \State $k, optimal\_solution\leftarrow$ D\textsc{imension}R\textsc{eduction}($\mathcal{C},M,\lambda$)
          \State $\varepsilon^{\varLambda[k]}\leftarrow optimal\_solution$ 
          \Comment{$\varLambda[k]$: the $k$-th element of $\varLambda$}
          \State $\mathcal{C}\leftarrow$ L\textsc{ocal}R\textsc{ecombination}($\mathcal{C},k,optimal\_solution,\tau$)
          \State delete $k$-dimensional data of $\mathcal{C}$
          \State $\varLambda\leftarrow\varLambda \backslash \{k\}$ \Comment{remove $k$ from $\varLambda$}
        \EndWhile
      \State $\mathbb{T}\leftarrow\mathbb{T}\bigcup\{\varepsilon\}$
      \EndWhile
      \For{$k=1\rightarrow |\mathbb{T}|$}
        \State $\mathbb{T}[k]\leftarrow\mathcal{F}(\mathbb{T}[k])$
      \EndFor
      \State \Return $\mathbb{T}$
    \EndFunction
  \end{algorithmic}
}
\end{algorithm}

\subsection{Dimension Reduction}
As PRT generates a new test case dimension by dimension, dimension reduction compares the largest intervals of every dimension in the candidate set $\mathcal{C}$, generates the corresponding value. \add{More specifically, given a $M$-dimensional test case set $\mathcal{C}=\{\varepsilon_1, \varepsilon_2, \cdots, \varepsilon_n\}$, the largest interval of $i$-th dimension is defined as $E_i$:}

\vspace{-0.2cm}
\begin{equation}
  \add{
  \begin{array}{ccc}
    \underset{x \in \text{InD}^i}{\max} & E_i & \\
    \text{s.t.} & E_i \le |x-\varepsilon^i_k|,& k = 1,2,\dots,n 
  \end{array}
  }
\end{equation}

\noindent \add{where $1\le i \le M$, $InD^i$ is the $i$-dimensional input domain, $\varepsilon^i_k$ is the $i$-dimensional value of $\varepsilon_k$. And the optimal solution $x$ is referred to the corresponding solution for $E_i$.}

\noindent\textbf{Idea.} Given $n$ points in high-dimensional domain, finding the farthest point from them is an NP-hard (Nondeterministic Polynomial time) problem. However, the problem is easy to be solved in one-dimensional domain. 


\modify{Fig.~\ref{fig:ruler} gives an example in one-dimensional domain}{For example}, assuming the input domain is $[0,1]$\modify{. We}{, we} have four existing points (colored blue): $p_1=0,p_2=0.1,p_3=0.3,p_4=0.7$. We can compute their interval as Eq.~\ref{eq:diff}. 

\vspace{-0.2cm}
\begin{equation}
  (p_2,p_3,p_4) - (p_1,p_2,p_3) = (0.1,0.2,0.4)
  \label{eq:diff}
\end{equation}

It is easy to find a large interval between $p_3$ and $p_4$. The corresponding solution is $p=(p_3+p_4)/2=0.5$ and the interval is $(p_4-p_3)/2=0.2$. However, by taking the infimum and the supremum into account, we find that $p=1$ is the optimal solution because of its larger interval $1-p_4=0.3>(p_4-p_3)/2=0.2$. So the new generated point for this round is $p=1$ (colored green) and the maximal interval is $0.3$. Similarly, we can get the next generated point $p=0.5$ (colored orange) and its corresponding interval $0.2$. 

\modify{Furthermore, we expect the sparsest point is not a deterministic but a more likely solution, because we are not sure whether it is the best choice.}{Furthermore, the sparsest point is treated as a probabilistic rather than deterministic solution, as it may not always be the optimal choice.} We design a probabilistic function as Eq.~\ref{eq:maxprob}, where $\modify{a}{inf}$ and $\modify{b}{sup}$ are the infimum and the supremum respectively, $\lambda\ge 1$ is a hyperparameter, and $\theta\in[0,1]$ follows a uniform distribution. Eq.~\ref{eq:maxprob} makes the generated point more likely close to $\modify{(a+b)/2}{(inf+sup)/2}$ and $\lambda$ controls the probability. The larger $\lambda$ is, the more likely $\modify{f(\theta|a,b,\lambda)}{f(\theta\,\vert\, inf,sup,\lambda)}$ is close to $\modify{(a+b)/2}{(inf+sup)/2}$. When $\lambda=1$, $\modify{f(\theta|a,b,\lambda)}{f(\theta\,\vert\, inf,sup,\lambda)}$ follows a uniform distribution. 

\vspace{-0.2cm}
\begin{equation}
  \label{eq:maxprob}
  \modifycolor{
  f(\theta\,\vert\, inf,sup,\lambda) = \left\{\begin{array}{ll}
    \frac{inf+sup}{2}-(1-2\theta)^\lambda\cdot\frac{sup-inf}{2}, & 0\le\theta\le 0.5 \\
    \frac{inf+sup}{2}+(2\theta-1)^\lambda\cdot\frac{sup-inf}{2}, & 0.5<\theta\le 1
  \end{array}\right.
  }
\end{equation}

\noindent\add{Here $(inf+sup)/2$ is the middle point. As $\theta$ follows a uniform distribution $U(0,1)$, the generated point moves left when $\theta<0.5$, and moves right when $\theta>0.5$. $|1-2\theta|$ is an item ranging from $0$ to $1$. The larger $\lambda$, the smaller $|1-2\theta|^\lambda$ and the more likely $f(\theta\,\vert\, inf,sup,\lambda)$ close to the middle point. We refer $\lambda$ as the confidence level. It is correlated with the probability function of new generating points. More specifically, in the case of left boundary, $\theta$ is required to follow a uniform distribution $U(0.5,1)$. The larger $\lambda$, the more likely new points close to the boundary. And so like for the right boundary.}

\begin{algorithm}
  \caption{Dimension Reduction}
  \label{algo:dr}
  \setlength{\baselineskip}{1pt}
  \begin{algorithmic}[1]
    \Require test case set $\mathcal{C}$, dimension $M$, confidence $\lambda$
    \Ensure the sparsest dimension $k$, the corresponding solution $value$
    \Function{DimensionReduction}{$\mathcal{C},M,\lambda$}
      \State $c_1, c_2, \cdots, c_n \leftarrow$ all test cases in $\mathcal{C}$
      \State $intervals \leftarrow \emptyset$
      \State $solutions \leftarrow \emptyset$
      \For{$i = 1 \to M$}
        \State $p_1,p_2,\cdots,p_n\leftarrow sort(c^i_1, c^i_2, \cdots, c^i_n)$
        \Comment{in ascending order}
        \If{$n==1$}
          \State $max\_interval\leftarrow 0$
        \Else
          \For{$j=1\rightarrow n-1$}
            \State $d_j\leftarrow p_{j+1}-p_j$
          \EndFor
          \State $k\leftarrow \arg\max_j(d_j)$
          \State $max\_interval\leftarrow (p_{k+1}-p_k)/2$
          \State $\theta\leftarrow UniformSample(0,1)$
          \Comment Sample a value from a uniform distribution $[0,1]$
          \State $solution\leftarrow f(\theta\, |\, p_k,p_{k+1},\lambda)$
        \EndIf
        \If{$max\_interval < p_1 - inf^i$}
        \Comment $inf^i:$ the infimum of the $i$-th dimension
          \State $max\_interval\leftarrow p_1 - inf^i$
          \State $\theta\leftarrow UniformSample(0.5,1)$
          \State $solution\leftarrow f(\theta\, |\, 2\cdot inf^i-p_1,p_1,\lambda)$
        \EndIf 
        \If{$max\_interval< sup^i - p_n$}
        \Comment $sup^i:$ the supremum of the $i$-th dimension
          \State $max\_interval\leftarrow sup^i - p_n$
          \State $\theta\leftarrow UniformSample(0,0.5)$
          \State $solution\leftarrow f(\theta\, |\, p_n,2\cdot sup^i-p_n,\lambda)$
        \EndIf 
        \State $intervals\leftarrow intervals\bigcup\{max\_interval/(sup^i-inf^i)\}$
        \State $solutions\leftarrow solutions\bigcup\{solution\}$
      \EndFor
      \State $k\leftarrow\arg\max(intervals)$
      \State $value\leftarrow solutions[k]$
      \State \Return $k,value$
    \EndFunction
  \end{algorithmic}
\end{algorithm}

\noindent\textbf{Design.} Algo.~\ref{algo:dr} formulates the workflow to find the sparsest dimension and the corresponding value. Lines 5--30 record the maximal interval and the corresponding solution dimension by dimension. Lines 7--8 aim to only consider the boundary (lines 18--27). Lines 10--12 compute the intervals, and lines 13--14 record the largest one. Line 15 samples a value from the uniform distribution between $0$ and $1$. Line 16 computes the generated value according to Eq.~\ref{eq:maxprob}. Lines 18--27 take a similar method to deal with the boundary case, where $inf^i$ and $sup^i$ are the infimum and the supremum of $i$-dimensional input domain. Lines 28--29 record the analytical result of the $i$-th dimension after normalization. Line 31 selects the dimension with the maximal interval. Line 32 obtains the solution by the dimension number. 

\vspace{-0.3cm}
\subsection{Local Recombination}

\add{\textbf{Idea.} Algo.~\ref{algo:dr} aims to identify the largest interval containing a given number of points across different dimensions. However, this task becomes challenging when the number of points is large and the dimensionality is high. Moreover, Algo.~\ref{algo:dr} fails to preserve the relationships among test cases across dimensions.} \modify{Two points' distance may be short in one dimension but long in another dimension.}{Therefore, we aim to reconstruct the candidate set $\mathcal{C}$. And we regard that the generation process should focus on points close to the newly generated $solution$.} \modify{Assuming that we are generating a new test case $\varepsilon$ from a set $\mathbb{T}$, local recombination makes the generating process focus on the similar test cases in $\mathbb{T}$, whose distance from $\varepsilon$ is lower than $\tau$ along all generated dimensions.}{} 

\modifycolor{
\noindent\textbf{Design.} Eq.~\ref{eq:lr} defines the local recombination strategy, where $k_1,k_2,\cdots,k_m$ are the generated dimensions of $\varepsilon$, $Q_\tau$ denotes $\tau$-quantile. 

\vspace{-0.3cm}
\begin{equation}
  \mathcal{C}' = Recombine(\mathbb{T}|\varepsilon,\tau) = \bigcap_{i=1}^{m} \left\{ x \in \mathbb{T} \;\middle|\; |x^{k_i} - \varepsilon^{k_i}| \le Q_\tau\!\left( \{\, |y^{k_i} - \varepsilon^{k_i}| : y \in \mathbb{T} \,\} \right) 
\right\}
  \label{eq:lr}
\end{equation}

If $k_{m+1}$ is newly generated dimension and $\mathcal{C}$ is the last recombination candidate set, Eq.~\ref{eq:lr} can be simplifed as only computing the $k_{m+1}$-th dimension:

\vspace{-0.3cm}
\begin{equation}
  \mathcal{C}' = Recombine(\mathcal{C}|\varepsilon,\tau) = \left\{ x \in \mathcal{C} \;\middle|\; |x^{k_{m+1}} - \varepsilon^{k_{m+1}}| \le Q_\tau\!\left( \{\, |y^{k_{m+1}} - \varepsilon^{k_{m+1}}| : y \in \mathcal{C} \,\} \right) 
\right\}
  \label{eq:slr}
\end{equation}

\begin{algorithm}
  \caption{Local Recombination}
  \label{algo:lr}
  \setlength{\baselineskip}{1pt}
  \begin{algorithmic}[1]
    \Require test case set $\mathcal{C}$, target dimension number $k$ and new value $solution$, resolution $\tau$
    \Ensure the recombined test case set $\mathcal{C}'$
    \Function{LocalRecombination}{$\mathcal{C},k,solution,\tau$}
      \State $Dist\leftarrow\emptyset$
      \For{$c \in \mathcal{C}$}
        \State $Dist\leftarrow Dist \bigcup \{|c^k-solution|\}$
      \EndFor
      \State $q_\tau \leftarrow Q_\tau(Dist)$
      \State $\mathcal{C}'\leftarrow \left\{c \in \mathcal{C} \;\middle|\; |c^k-solution| \le q_\tau\right\}$
      \State \Return $\mathcal{C}'$
    \EndFunction
  \end{algorithmic}
\end{algorithm}

Algo.~\ref{algo:lr} formulates the local recombination process. Line 6 obtains the $\tau$-quantile of the distance set. A simple method is to sort the points and obtain the $\tau|Dist|$-th point, whose time complexity is $O(n\log n)$. Here we leverage Introselect algorithm to reduce the time complexity to $O(n)$. 
}

\add{There is a question on how to select the hyperparameter $\tau$. From Algo.~\ref{algo:lr}, we can get $|\mathcal{C}|= \lceil\tau|\mathcal{C}|'\rceil \thickapprox \tau|\mathcal{C}|'$. Initially in the inner \textbf{while} loop of Algo.~\ref{algo:prt}, there are $|\mathbb{T}|$ points and the iteration will have $M$ times. By designing $N\cdot\tau^M=1$, we have a recommended setting:

\begin{equation}
  \tau=N^{-\frac{1}{M}}
\end{equation}
}

\add{\subsection{Boundary Priority}
In this section, we explain why PRT prioritizes boundary regions by default.

First, Algo.~\ref{algo:dr} assigns a boundary interval that is twice as large as that of a middle region, prioritizing generating boundary points. For example, suppose the input domain is $[0,1]$ and the existing points are ${0.3, 0.8}$. The interval corresponding to the candidate solution at 0.55 is only 0.25, whereas the interval at the boundary point 0 is 0.3. Thus, the next generated point will be 0.

Moreover, Algo.~\ref{algo:lr} further reinforces boundary exploration across dimensions. For instance, assume the existing points lie on the boundary in the $i$-th dimension but in the middle region of the $j$-th dimension. Although the new point may be generated in the middle region along the $i$-th dimension, it will be positioned on the boundary along the $j$-th dimension.
}

\subsection{Linear Mapping}
\label{linearmapping}
We demonstrate that PRT generates evenly distributed test cases but prioritizes testing the boundary by default ($\mathcal{F}(t)=t$). However, sometimes we may expect PRT to prioritize other regions. Here we give an example of prioritizing the middle domain. Eq.~\ref{eq:mapping} assumes $t$ is a scalar value with infimum $\modify{a}{inf}$ and supermum $\modify{b}{sup}$ and maps the boundary to the middle $\modify{(a+b)/2}{(inf+sup)/2}$. \modify{For high dimensional data, just do so for every dimension.}{For high-dimensional data, this process is repeated for every dimension.}

\begin{equation}
  \label{eq:mapping}
  \mathcal{F}_1(t)=\left\{\begin{array}{ll}
    \modify{\frac{3a+b}{2}-t}{\frac{3inf+sup}{2}-t}, & \modify{a}{inf}\le t \le \modify{\frac{a+b}{2}}{\frac{inf+sup}{2}} \\
    \modify{\frac{a+3b}{2}-t}{\frac{inf+3sup}{2}-t}, & \modify{\frac{a+b}{2}}{\frac{inf+sup}{2}}< t \le \modify{b}{sup} \\
  \end{array}\right.
\end{equation}

Note that we make a linear transformation to ensure the evenly distributed property. Other transformations can also be adopted for specific purposes. 

\subsection{Realization}

If we follow the steps of Alg.~\ref{algo:dr} and Alg.~\ref{algo:lr}, \modify{the program is really slow}{the time complexity is $O(MN^2\log N)$}. We use some data structure techniques to accelerate the program\modify{, where the key point is to store the sorted results and indexes}{}.

Assume that we want to generate $N$ $M$-dimensional test cases and we already have $k$ test cases during the \textbf{while} loop in Alg.~\ref{algo:prt}. \modify{Firstly, use a $M\times N$ matrix $D$ to store test cases.}{First, an $M \times N$ matrix $D$ is used to store test cases.} $i$-th row records the $i$-dimensional data in an ascending order, despite the relationship between different dimensions. That makes the time complexity of Alg.~\ref{algo:dr} line 6 $O(1)$ and that of the whole Alg.~\ref{algo:dr} $O(Mk)$. Secondly, use a $M\times k$ matrix $\alpha$ to record which number of test case $D(i,j)$ is from, and use a $k\times M$ matrix $\beta$ to record which index of $D$ the $k$-th test case is in. They satisfy $\alpha(m,\beta(j,m))=j,\forall 1\le j\le k, 1\le m\le M$. That makes the time complexity of the whole Alg.~\ref{algo:lr} $O(\tau k(M-1))$. At the next iteration, the number of test cases approximate $\tau k$ and the time complexity of Algo.~\ref{algo:dr} and Algo.~\ref{algo:lr} will be $O(\tau k(M-1))$ and $O(\tau^2 k (M-2))$ respectively. \modify{Both $O(Mk)$ and $O(M\tau k\log_2 (\tau k))$ are the time complexity of line \modify{8}{9} and line \modify{10}{11} in Alg.~\ref{algo:prt} at the first round. Noticing that the dimension reduction method always generates the sparsest point, the number of points will decrease with an exponential speed $\tau$ due to the local recombination method. So in the next dimension's generating round, the time complexity will be changed to $O(M\tau k)$ and $O(M\tau^2 k \log_2(\tau^2 k))$. Then we can get the time complexity of Lines \modify{7--12}{8--14} in Algo.~\ref{algo:prt} as the maximum of Eq.~\ref{eq:o1} and Eq.~\ref{eq:o2}. By traversing $k$ from $1$ to $N$ we can get the result $O(MN^2\frac{1-\tau^M}{1-\tau}\cdot\max(1,\tau \log_2 (\tau N) ))$. If we regard $\tau$ as a constant and $\log_2 (\tau N) $ as a small number, the time complexity approximates $O(MN^2)$.}{As the time complexity of Algo.~\ref{algo:lr} is lower than that of Algo.~\ref{algo:dr}, the time complexity of Lines 8--14 in Algo.~\ref{algo:prt} is $O\left(\tau^t k(M-t)\right)$, where $k$ and $t$ implies the $k$-th outer \textbf{while} loop and the $t$-th inner \textbf{while} loop. By traversing $k$ from $1$ to $N$ we can get the result $O(MN^2)$.}

\modifycolor{
\begin{equation}
  \label{eq:o}
  \sum_{k=2}^{N} \sum_{t=0}^{M-1} O\left(\tau^t k(M-t)\right) \big|_{\tau=k^{-\frac{1}{M}}}
  \approx \sum_{k=2}^{N} O(M k) 
  = O(MN^2)
\end{equation}
}

\section{Experimental Setup}

\subsection{Baselines}
PRT is a black-box setting method and can be applied to test any type of DRL agents. Our baselines include SOTA fuzzing methods MDPFuzz \cite{mdpfuzz} and CureFuzz \cite{curefuzz}, search-based method QD \cite{qd}, generative-based method G-Model \cite{gmodel} as well as random testing (RT). For QD, we use the Map-Elites algorithmic version as it performs the best in most cases analyzed by the authors \cite{qd}. Note that though these methods are all black-box, most of them need preparation before testing. Besides, the information used for testing also varies. 

\begin{table}[h!]
  \begin{center}
    \caption{Applicable conditions of different methods.}
    \label{tab:reliance}
    \setlength{\extrarowheight}{1pt}
    \begin{tabularx}{1.0\textwidth}{>{\centering\arraybackslash}p{2.1cm} >{\centering\arraybackslash}p{2cm} >{\centering\arraybackslash}X >{\centering\arraybackslash}X | >{\centering\arraybackslash}p{1.2cm} >{\centering\arraybackslash}p{1.2cm} >{\centering\arraybackslash}p{1.2cm}}
      \hline
      & Failure & \multirow{2}{*}{Execution} & Model & \multirow{2}{*}{State} & \multirow{2}{*}{Action} & \multirow{2}{*}{Reward} \\
      & Insights & & Training &  &  &  \\
      \hline
      \textbf{PRT} (ours) & $\surd$ & $\times$ & $\times$ & $\times$ & $\times$ & $\times$ \\
      MDPFuzz \cite{mdpfuzz} & $\times$ & $\surd$ & $\surd$ & $\surd$ & $\times$ & $\surd$ \\
      CureFuzz \cite{curefuzz} & $\times$ & $\surd$ & $\surd$ & $\surd$ & $\times$ & $\surd$ \\
      G-Model \cite{gmodel} & $\times$ & $\times$ & $\surd$ & $\times$ & $\times$ & $\times$ \\
      QD \cite{qd} & $\times$ & $\surd$ & $\times$ & $\surd$ & $\surd$ & $\surd$ \\
      RT & $\times$ & $\times$ & $\times$ & $\times$ & $\times$ & $\times$ \\
      \hline
    \end{tabularx}
  \end{center}
\end{table}

Tab.~\ref{tab:reliance} summarizes their differences. PRT needs failure insights about the task. MDPFuzz, CureFuzz and QD need to execute a number of test cases before testing. MDPFuzz, CureFuzz and G-Model will train machine learning models before testing. Furthermore, these methods access different information during testing. MDPFuzz and CureFuzz require the executing data, including state and reward, while QD requires state, action and reward.

For MDPFuzz \cite{mdpfuzz} and CureFuzz \cite{curefuzz}, which require executing test cases before testing, we randomly generate 2000 passing test cases, where we choose 2000 to give the method enough information and the passing condition is their inner algorithmic step. Though QD \cite{qd} needs execution before testing, it incorporates the execution into the testing cost. 

\subsection{Testing Subjects}
\label{testingsubjects}

We select the three most popular environments in Tab.~\ref{tab:expertknowledge} \add{and one high-dimensional environment} for testing, which are all developed by OpenAI in Gymnasium platform \cite{gym}. The three most popular environments are also chosen because of the different reward patterns (RQ1 in Sec.~\ref{rq1}, Fig.~\ref{fig:rewardpattern}). We introduce the environments' tasks, the test case definitions and the test oracles as follows. To comprehensively test different types of agents, we select the discrete version of Cart Pole and the continuous versions of Lunar Lander and Mountain Car. 


\vspace{-0.2cm}
\begin{figure}[htbp]
  \centering
  \subfloat[Cart Pole\label{fig:cartpole}]{
    \includegraphics[width=0.24\textwidth]{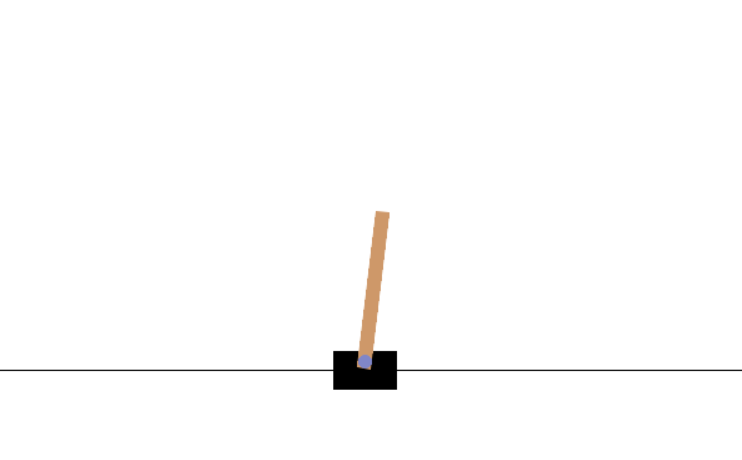}
  }
  \subfloat[Lunar Lander\label{fig:lunarlander}]{
    \includegraphics[width=0.24\textwidth]{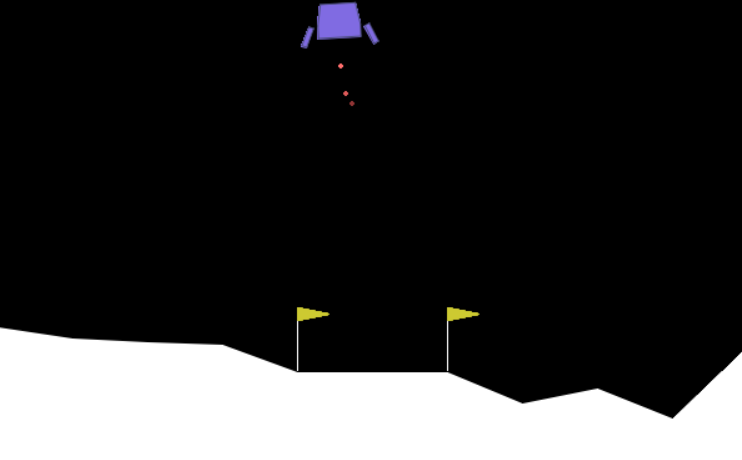}
  }
  \subfloat[Mountain Car\label{fig:mountaincar}]{
    \includegraphics[width=0.24\textwidth]{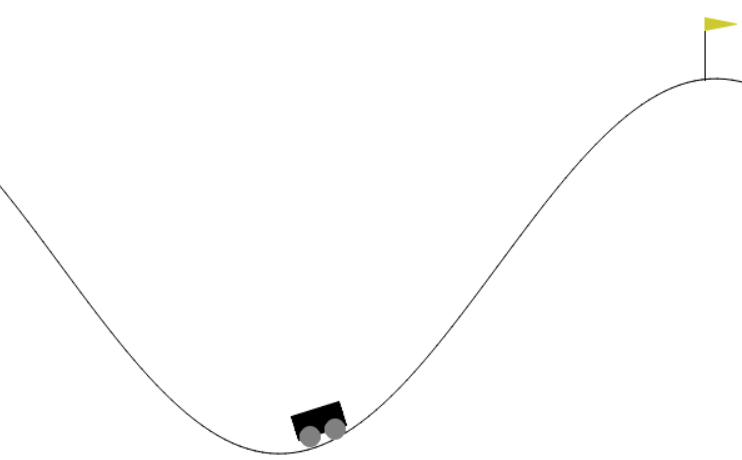}
  }
  \subfloat[Humanoid\label{fig:humanoid}]{
    \includegraphics[width=0.24\textwidth]{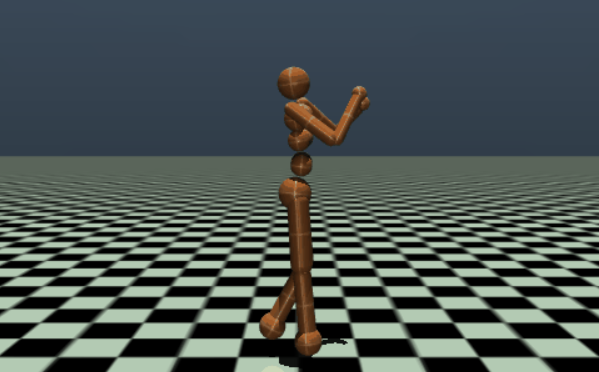}
  }
  \caption{Different environment examples}
  \label{fig:environment}
\end{figure}

\vspace{-0.4cm}
\subsubsection{Cart Pole} This environment corresponds to the version of the cart-pole problem described by Barto et al. \cite{cartpole}. In Fig.~\ref{fig:cartpole}, a pole is attached by an unactuated joint to a cart, which moves along a frictionless track. The test case is defined as its four initial states $(x,\dot{x},\theta,\dot{\theta})$: the cart's position $x\in[-0.9,0.9]$, the cart's velocity $\dot{x}\in[-0.3,0.3]$, the pole's angle $\theta\in[-\frac{\pi}{60},\frac{\pi}{60}]$ and the pole's angular velocity $\dot{\theta}\in[-\frac{\pi}{90},\frac{\pi}{90}]$. The agent can apply either a left or a right force to the cart. Once the pole's angle is not within $\pm12^\circ$ or the cart's position is not within $\pm2.4$, the task fails. If the agent fails at $t$ step, it will have a reward $t$. If the system keeps balance during 500 steps, we will terminate testing, give a reward 500 and regard the test case as passing.

\subsubsection{Lunar Lander} 
This is a classic rocket trajectory optimization problem. In Fig.~\ref{fig:lunarlander}, a lander descends from the middle sky towards a designated landing pad on the lunar surface. The test case is defined as the initial force $(f_x,f_y)$ applied to the lander along $x$ direction $f_x\in[-1500,1500]$ and $y$ direction $f_y\in[-1500,1500]$. The agent can make one of four discrete actions: do nothing, fire the left orientation engine, fire the right orientation engine, or fire the main engine. A task fails if the lander crashes (e.g., outside the landing pad or with excessive speed), while it succeeds if the lander safely lands on the middle pad. The reward is calculated according to the distance to the landing pad, the speed, the angle, and the leg contacts. A successful landing within the designated area yields a high reward of 100, while crashing results in a large negative reward of -100.  

\subsubsection{Mountain Car} 
This environment first appeared in Andrew Moore's PhD thesis (1990) \cite{moutaincar}. In Fig.~\ref{fig:mountaincar}, a car is situated between two hills and must build momentum to reach the flag positioned at the top of the right hill. The test case is defined by its two initial states $(x,v)$: the car's horizontal position $x\in[-1,0]$ and velocity $v\in[-0.05,0.05]$, where the position of the valley is $-\frac{\pi}{6}\approx -0.52$ and the position of the flag is $0.5$. The agent can apply one of three discrete actions: accelerate left, accelerate right, or do nothing. The agent receives a constant reward of $-1$ per step until termination. If the car reaches the flag at time step $t$, the task succeeds with a cumulative reward of $-t$; but if $t>200$, the test case is regarded as a failure.  

\add{\subsubsection{Humanoid-v4}
This environment is a 3D bipedal robot simulation based on the work of Tassa et al. \cite{humanoid}. In Fig.~\ref{fig:humanoid}, a humanoid robot with a torso, a pair of arms and legs (each leg has three parts, each arm has two parts) is designed to simulate human locomotion. The robot starts in a standing pose, and the test case is defined as the bias added to the standing pose. The input domain is $[-0.01,0.01]^{47}$, including 24 position values and 23 velocity values. The agent controls the robot by applying torques to its 17 hinge joints, including 3 for the abdomen, 3 for each hips, 1 for each knees, 2 for each arm's shoulder, and 1 for each elbow, totally $3+(3+1+2+1)\times 2=17$. The goal is to make the robot walk forward as fast as possible without falling. The task fails if the robot falls (the torso height drops below a threshold). At each time step, the agent receives a reward comprised of a forward velocity bonus, a healthy reward for staying upright, and penalties for excessive control forces. If the agent survives until the time horizon (1000 steps), the test case is regarded as passing.}

\noindent\textbf{Validity of test cases.} \add{We regard that a test case $\varepsilon$ is valid if and only if there exists a feasible action sequence enabling the agent to pass the test case.} \modify{We have verified that the most difficult task in our designed input domain can successfully be solved, which means the failures are due to the wrong decision by the agent rather than the design of test cases.}{We have verified that the most difficult test cases (i.e., the most difficult tasks of specific environments) within the designed input domain can successfully be solved, which implies all failures are caused by incorrect software logic (i.e., incorrect agent actions) rather than flaws in the input domain design.} \modify{Though all test cases in our input domain for the three environments are valid, we still keep the interface for checking whether the generated test case is valid in the realization of every method. Once an invalid test case is generated, all methods (we realize all methods in Tab.~\ref{tab:reliance}) take a general solution: drop the test case itself and generate another.}{In addition, as we change the initial state bound of Cart Pole, Lunar Lander and Mountain Car, we upload three oracle agents to our open-source repository for verifying that all failures can be solved.}

\noindent\textbf{Task-induced failure insights.} In the Cart Pole environment (Fig.~\ref{fig:cartpole}), the agent must balance the cart and the pole with $(x,\dot{x},\theta,\dot{\theta})$ around $(0,0,0,0)$, so the most difficult task corresponds to the boundary $(\pm 0.9,\pm 0.3,\pm \frac{\pi}{60},\pm \frac{\pi}{90})$. In the Lunar Lander environment (Fig.~\ref{fig:lunarlander}), the lander is required to safely land on the middle pad, so the most difficult task also corresponds to the boundary $(\pm 1500, \pm 1500)$. In the Mountain Car environment (Fig.~\ref{fig:mountaincar}), the car must climb the right hill to reach the flag within specific steps, so the most difficult task corresponds to the middle point $(-0.5,0)$. \add{In the Humanoid environment (Fig.~\ref{fig:humanoid}), the robot needs to maintain the standing pose, so the most difficult task corresponds to the boundary $(\pm 0.01, \pm 0.01, \cdots, \pm 0.01)$.}

\subsection{Hyperparameters}
\label{hyperparameter}
\textbf{Mapping $\mathcal{F}$.} For \modify{Cart Pole and Lunar Lander}{Cart Pole, Lunar Lander and Humanoid}, the mapping $\mathcal{F}$ does not change anything $\mathcal{F}(t)=t$. For the Mountain Car environment, we shift the boundary to the middle with Eq.~\ref{eq:mapping} $\mathcal{F}_1$. 


\noindent\textbf{Confidence $\lambda$.} We set $\lambda=20$. Under this setting, assuming the parameter $\theta$ follows a uniform distribution $\theta \sim U(0,1)$, the deviation term $\modify{2^{\lambda}(0.5-\theta)^{\lambda}}{(1-2\theta)^{\lambda}}$ in Eq.~\ref{eq:maxprob} exceeds $10\%$ with probability $10.9\%$, and exceeds $2\%$ with probability $17.8\%$.

\subsection{Implementation}
\noindent\textbf{Stochasticity of test case execution.} Stochasticity is a common challenge in testing DRL agents. To minimize its impact on the outputs, we generate test cases using different random seeds but execute all test cases under a fixed random seed\add{, which means the test case generation is non-deterministic but the test case execution is deterministic}.

\noindent\textbf{Experimental platform.} Our experimental platform is 24.04.1-Ubuntu, with hardwares consisting of an AMD Ryzen 9 9950x CPU and two \SI{32}{GB} \SI{5200}{Hz} RAMs. We do not use any GPU for testing. 

\section{Evaluation}
In this section, we present three research questions (RQs) to study the effectiveness, efficiency and generalization of PRT. 

\emph{RQ1: \modify{Can PRT effectively and efficiently find failures?}{How effective and efficient is PRT for detecting failures?}} 

We evaluate PRT against SOTA fuzzing, search-based, and generative methods on DRL agents with \modify{random failure rates below 1\%}{failure rates below 1\% measured by random testing under 100,000 tests}. The comparison focuses on the testing cost required to detect the first failure. In addition, we examine the conditions under which PRT outperforms the baselines, as well as those in which the baselines demonstrate superior performance.

\emph{RQ2: How diverse are test cases generated by different methods?} 

\modify{We evaluate the diversity by the coverage of test cases. A larger coverage can help to miss fewer failures. Moreover, we visualize the distribution of test cases to enable deeper insights into their exploration strategies.}
{As we claim that PRT preserves the diversity of generated test cases while enabling prioritization, RQ2 evaluates the extent of this diversity, particularly in comparison with the baselines. In addition, we visualize the distributions of all generated test cases to provide further insights into the strengths and limitations of different algorithms beyond scalar diversity metrics.}

\emph{RQ3: What are the failure patterns of different agents, and how does PRT address them?} 

\modify{Since task-induced failure insights originate from the environment and are only indirectly related to specific agents, we aim to verify that such insights are environment-dependent but largely agent-independent.}{As a failure-based testing method, the effectiveness of PRT is strongly correlated with the failure pattern of testing subjects.} To this end, we train agents with different algorithms and analyze their failure patterns. We then investigate why PRT can effectively detect failures across different DRL agents.

\subsection{Diversity Metric}
\label{sec:diversity}
\add{We quantify the diversity of generated test cases using differential \textbf{entropy}, which measures the dispersion of samples in continuous state spaces. Given a continuous random variable $X \in \mathbb{R}^d$ with probability density function $p(x)$, its differential entropy is defined as:

\vspace{-0.2cm}
\begin{equation}
  H(X) = -\int p(x)\log p(x)\, dx
\end{equation}

Differential entropy characterizes the effective volume occupied by a distribution and increases as samples become more dispersed within the same support \cite{entropy=effectivevolume}. Under identical dimensionality and normalization, higher entropy indicates broader exploration of the state space rather than closeness to a specific target distribution.

Since the underlying distribution of test cases is unknown and may exhibit complex, non-Gaussian structures, we adopt a non-parametric k-nearest neighbor (kNN) entropy estimator. Specifically, we use the estimator proposed by Kozachenko and Leonenko \cite{sampleestimate}. Given a set of $n$ samples ${x_i}_{i=1}^n \subset \mathbb{R}^d$, the entropy is estimated as:

\vspace{-0.2cm}
\begin{equation}
  \label{eq:entropy}
  \hat{H}_k = \psi(n) - \psi(k) + \log V_d + \frac{d}{n}\sum_{i=1}^{n} \log \varepsilon_k(x_i)
\end{equation}

where $\psi(\cdot)$ denotes the digamma function, $V_d$ is the volume of the unit ball in $\mathbb{R}^d$, and $\varepsilon_k(x_i)$ is twice the Euclidean distance from $x_i$ to its $k$-th nearest neighbor. Unless otherwise stated, we fix the neighborhood size to $k = 10$ in all experiments, following common practice in non-parametric entropy estimation \cite{sampleestimate,beirlant1997nonparametric,NIPS2017_ef72d539}.

To ensure meaningful comparison, all entropy values are computed under identical state dimensionality, normalization, and neighborhood size, and are compared only within the same environment. Under these controlled conditions, the estimated entropy serves as a relative measure of sample dispersion, reflecting the extent to which an algorithm explores the continuous state space.}

\subsection{RQ1: Effectiveness and Efficiency for Detecting Failures}
\label{rq1}

\add{At the beginning, we construct some well-trained agents for testing. We employ PPO \cite{ppo} to train Cart Pole and Lunar Lander agents, TQC \cite{tqc} to train Mountain Car agents, and DSAC-T \cite{dsac} to train Humanoid agents. We repeat training and random testing for 100,000 times until two conditions are both satisfied: (1) there is at least one failure test case under 100,000 tests, and (2) the failures are less than 500. Specifically, the numbers of failures under 100,000 tests are 226, 164, 56, 112 for Cart Pole, Lunar Lander, Mountain Car and Humanoid respectively. We use the F-measure (the expected number of tests required to detect the first failure), proposed by Chen et al.~\cite{art}, as an effectiveness metric, as it directly reflects the failure-detection capability of a testing algorithm.}

\begin{table}[h!]
  \begin{center}
    \caption{Testing cost for finding the first failure \add{with corresponding ($p$-value, effect size)}}
    \label{tab:f1}
    \setlength{\extrarowheight}{1pt}
    \begin{threeparttable}
      \begin{tabularx}{\textwidth}{>{\centering\arraybackslash}X >{\centering\arraybackslash}X >{\centering\arraybackslash}X >{\centering\arraybackslash}X >{\centering\arraybackslash}X >{\centering\arraybackslash}X >{\centering\arraybackslash}X >{\centering\arraybackslash}X >{\centering\arraybackslash}X}
        \toprule
        & \multicolumn{2}{c}{Cart Pole} & \multicolumn{2}{c}{Lunar Lander} & \multicolumn{2}{c}{Mountain Car} & \multicolumn{2}{c}{Humanoid} \\
        & {\small Number} & {\small Time} & {\small Number} & {\small Time} & {\small Number} & {\small Time} & {\small Number} & {\small Time} \\
        \midrule
        \textbf{{\small PRT}} & \textcolor{myred}{62.65} & \textcolor{myred}{1.85} & \textcolor{myred}{70.15} & \textcolor{myred}{1.60} & \textcolor{myblue}{444.3} & \textcolor{myred}{2.20} & \textcolor{myred}{524.9} & \textcolor{myred}{165.9}\\[5pt]
        \multirow{2}{*}{{\small MDPFuzz}} & 2341.4 & 465.7 & 616.1 & 92.7 & \textcolor{myred}{96.0} & 4.68 & 1268.0\tnote{1} & 400.1\tnote{1} \\[-5pt]
        & {\footnotesize (0.00,0.98)} & {\footnotesize (0.00,1.00)} & {\footnotesize (0.00,0.92)} & {\footnotesize (0.00,0.98)} & {\footnotesize \textcolor{orange}{(1.00,0.19)}} & {\footnotesize (0.00,0.79)} & {\footnotesize (0.03,0.68)} & {\footnotesize (0.02,0.69)} \\[5pt]
        \multirow{2}{*}{{\small CureFuzz}} & 2182.6 & 68.4 & 613.5 & 15.1 & 1502.4 & 8.42 & \textcolor{myblue}{1116.6} & \textcolor{myblue}{352.4} \\[-5pt]
        & {\footnotesize (0.00,0.93)} & {\footnotesize (0.00,0.93)} & {\footnotesize (0.00,0.91)} & {\footnotesize (0.00,0.92)} & {\footnotesize (0.00,0.80)} & {\footnotesize (0.00,0.82)} & {\footnotesize \textcolor{orange}{(0.21,0.58)}} & {\footnotesize \textcolor{orange}{(0.20,0.58)}} \\[5pt]
        \multirow{2}{*}{{\small G-Model}} & 657.5 & 24.5 & 836.3 & 24.2 & 96192+\tnote{2} & 968+\tnote{2} & 4008.2 & 1310 \\[-5pt]
        & {\footnotesize (0.00,0.81)} & {\footnotesize (0.00,0.83)} & {\footnotesize (0.02,0.69)} & {\footnotesize (0.01,0.71)} & {\footnotesize (0.00,1.00)} & {\footnotesize (0.00,1.00)} & {\footnotesize (0.00,0.94)} & {\footnotesize (0.00,0.94)} \\[5pt]
        \multirow{2}{*}{{\small QD}} & 1616.5 & 48.6 & 566.9 & 13.2 & 656.1 & \textcolor{myblue}{4.23} & 1789.3 & 572.3 \\[-5pt]
        & {\footnotesize (0.00,0.89)} & {\footnotesize (0.00,0.89)} & {\footnotesize (0.00,0.88)} & {\footnotesize (0.00,0.88)} & {\footnotesize (0.01,0.74)} & {\footnotesize (0.00,0.78)} & {\footnotesize \textcolor{orange}{(0.17,0.59)}} & {\footnotesize \textcolor{orange}{(0.14,0.60)}} \\[5pt]
        \multirow{2}{*}{{\small RT}} & \textcolor{myblue}{491.5} & \textcolor{myblue}{15.2} & \textcolor{myblue}{532.8} & \textcolor{myblue}{12.7} & 2177.1 & 10.26 & 1261.3 & 393.2 \\[-5pt]
        & {\footnotesize (0.00,0.89)} & {\footnotesize (0.00,0.90)} & {\footnotesize (0.00,0.89)} & {\footnotesize (0.00,0.89)} & {\footnotesize (0.00,0.79)} & {\footnotesize (0.00,0.79)} & {\footnotesize (0.01,0.75)} & {\footnotesize (0.01,0.75)} \\
        \bottomrule
      \end{tabularx}
      \vspace{0.2cm}
      \footnotesize 
      \item The item ranking first is colored red, while the second is colored blue. \add{The statistics with $p$-value larger than 5\% is colored orange.} Number and Time mean the number of generated test cases and the corresponding time (in seconds, including algorithmic computation and environment simulation) for finding the first failure. 
      \add{\item [1] In the Humanoid environment, MDPFuzz requires approximately 553 seconds to analyze a single test case, implying that preparing 2,000 seeds would take more than 12 days before testing can even begin. Due to this prohibitive computational cost, we disable the freshness-guided component of MDPFuzz in the Humanoid environment.}
      \item [2] \modify{Across 20 experiments, G-Model succeeded in finding failures only once, at the 23,851st test case. In the remaining 19 experiments, testing was terminated after 100,000 trials without detecting any failures. Here, 96192 and 968 mean the minimal cost.}{Across 20 independent experiments, G-Model detected a failure only once, occurring at the 23,851st test case. In the remaining 19 experiments, testing was terminated after 100,000 trials without identifying any failures. Consequently, the minimum cost required for G-Model to detect a failure is (23,851+100,000*19)/20=96,192 in terms of test executions and 968 in terms of accumulated time cost.}
    \end{threeparttable}
  \end{center}
\end{table}

Under the low failure rate condition, we compare the testing cost of different methods to find the first failure test case. We consider the testing cost from two aspects: the number of generated test cases and the total testing time. For simulated programs, we usually want to find failures as fast as possible. However, for some practical applications, we may consider the cost of executing a test case, where the number of test cases is more important. For each experiment, we repeat it 20 times and report the average cost \add{together with the p-value and the effect size}. \add{For each experiment, we repeat the evaluation 20 times and report the average cost, together with the corresponding $p$-value and effect size. We employ Mann-Whitney U test to compute the $p$-value, under the one-sided assumption that the cost of the baselines is $lower$ than that of PRT. The effect size between the baselines and PRT is measured using the $\hat{A}_{21}$ statistic \cite{a12}.}

\begin{figure}[htbp]
  \centering
  \subfloat[Cart Pole\label{fig:cartpole_reward}]{
    \includegraphics[width=0.35\textwidth]{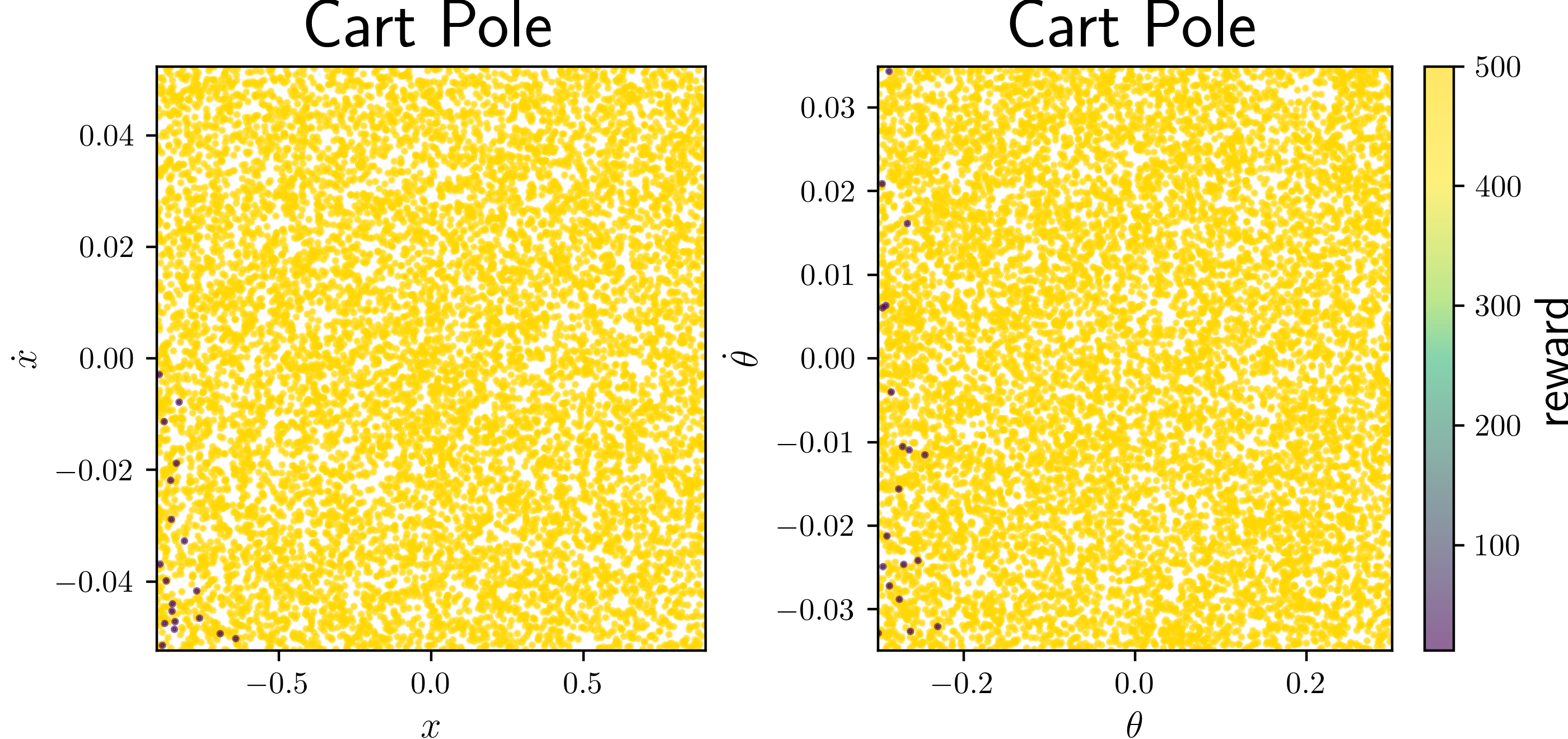}
  }
  \subfloat[Lunar Lander\label{fig:lunarlander_reward}]{
    \includegraphics[width=0.2\textwidth]{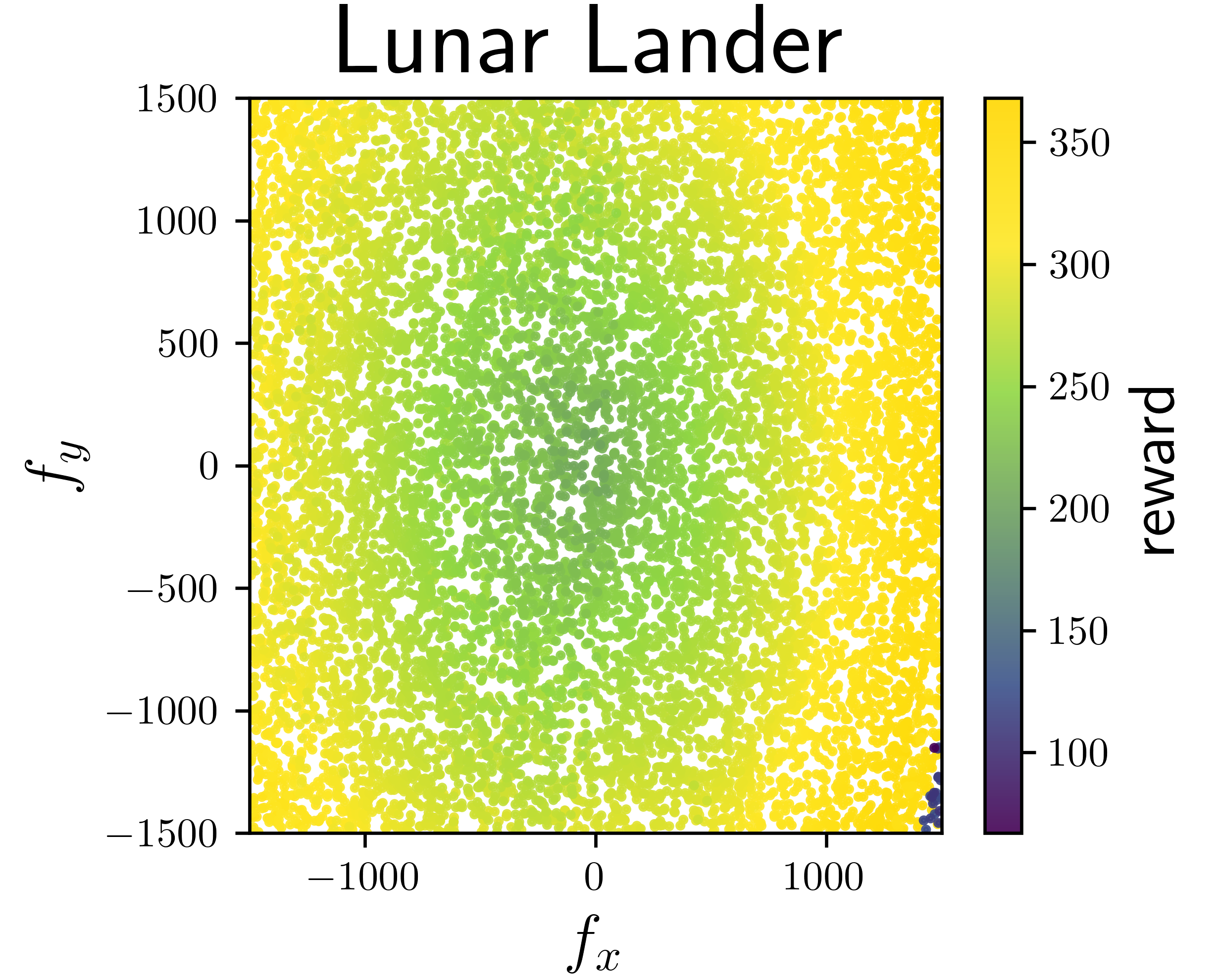}
  }
  \subfloat[Mountain Car\label{fig:mountaincar_reward}]{
    \includegraphics[width=0.2\textwidth]{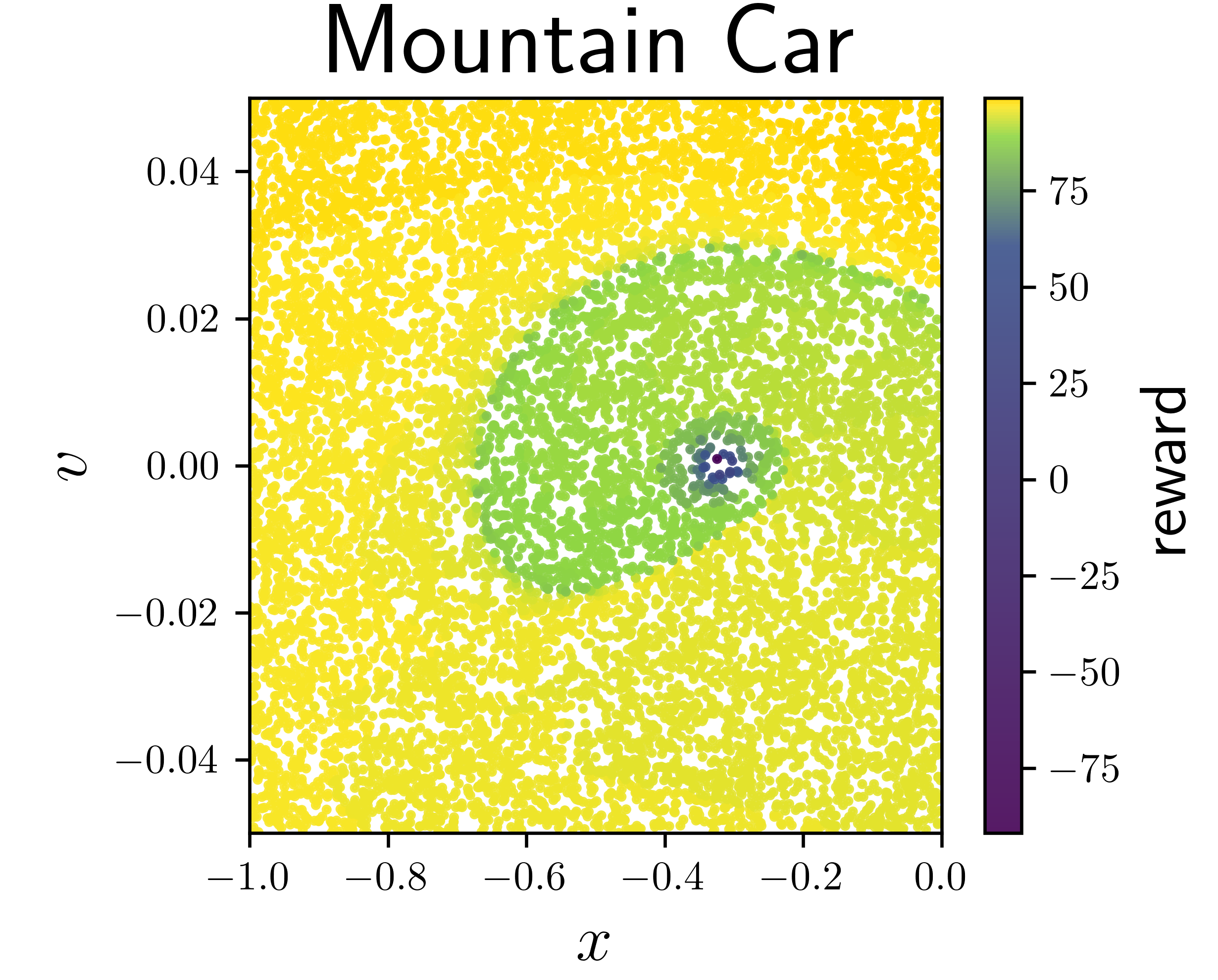}
  }
  \subfloat[Humanoid\label{fig:humanoid_reward}]{
    \includegraphics[width=0.2\textwidth]{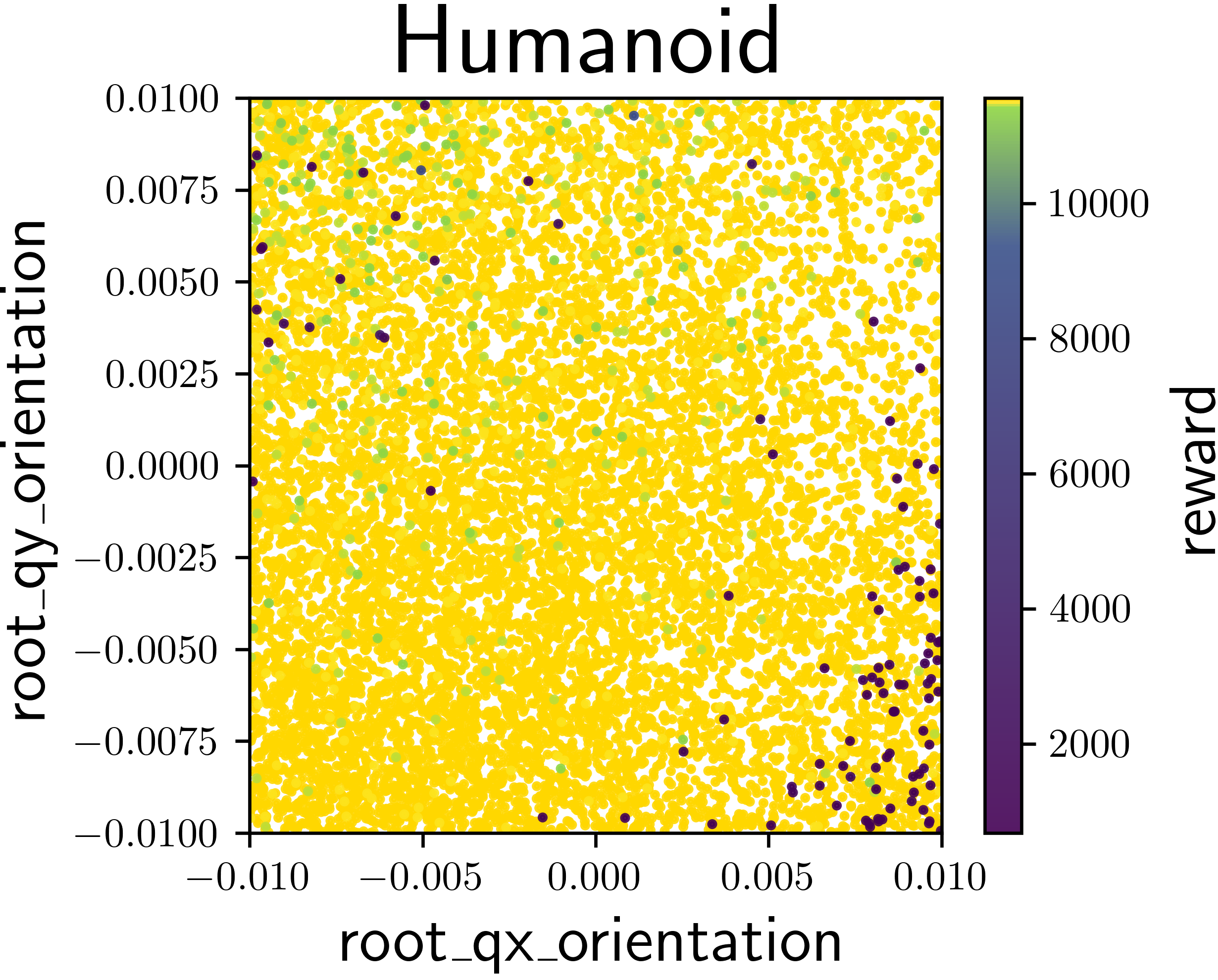}
  }
  \caption{Reward patterns of different environments}
  \label{fig:rewardpattern}
\end{figure}

Tab.~\ref{tab:f1} reports the average testing cost required by different methods to detect the first failure across various environments. In terms of average cost, PRT achieves the best performance in all cases, except for the number cost in Mountain Car. On one hand, the failure insights towards Mountain Car are deviated, where the failures are not precisely located in the central region. On the other hand, the reward pattern has strong guidance in this environment and MDPFuzz makes an aggressive strategy. Further analysis will be provided later in Fig.~\ref{fig:rewardpattern} and Fig.~\ref{fig:seedspattern}. Nevertheless, compared with RT, PRT consistently reduces the testing cost by at least 40\%, requiring significantly fewer test cases to detect the first failure. Moreover, PRT incurs low computational overhead, achieving the shortest time cost across all testing subjects. In Mountain Car, although the number cost of PRT is nearly six times that of MDPFuzz, it still detects the first failure faster.

\add{Tab.~\ref{tab:f1} also reports the corresponding $p$-values and effect sizes. Overall, PRT performs statistically better than RT across all environments. However, in some cases, the advantage of PRT is not statistically significant, particularly in the Humanoid environment. On one hand, the low average cost achieved by PRT can be attributed to its smaller maximal cost. In Humanoid, PRT's maximal number cost is 1461, whereas RT, MDPFuzz, CureFuzz, and QD exhibit maximal number costs of 3069, 5135, 7858, and 7992, respectively. As will be shown in RQ2, PRT achieves the highest diversity, which may contribute to its stability. On the other hand, the  statistical insignificance can be partly attributed to the high variance and limited repetitions in the experiments. The results indicate that, under limited experimental runs, some baselines can achieve competitive performance with PRT in specific scenarios, especially in the high-dimensional Humanoid environment.}

\modify{We observe that both MDPFuzz and QD achieve competitive results in Mountain Car. Notably, both MDPFuzz and QD benefit from reward guidance (primarily driven by rewards, though not exclusively).}{Focusing on the average cost, we observe that MDPFuzz achieves the best number cost in Mountain Car. Notice that MDPFuzz, CureFuzz, and QD all benefit from reward guidance (primarily, though not exclusively, driven by reward signals).} To better understand the underlying mechanisms, we further analyze the reward patterns.

\modify{For every environment, we execute 10,000 random test cases, obtain the corresponding rewards and then draw the heat map Fig.~\ref{fig:rewardpattern}.}{We draw the heat map as Fig.~\ref{fig:rewardpattern}. For Humanoid, we execute 100,000 random test cases and select 10,000 test cases with the lowest rewards. For the other environments, we execute 10,000 random test cases.} As the input domain of Cart Pole is four-dimensional, we present two plots (more details in Sec.~\ref{testingsubjects}). \add{For Humanoid, since the input domain has 47 dimensions, we enumerate all $\binom{47}{2} = 1081$ dimension pairs and select the pair with the lowest diversity among \textbf{failure test cases}, as computed by Eq.~\ref{eq:entropy}, which implies failures are clustered most at that projections.} The failure test cases achieve the lowest rewards and are colored dark purple. More specifically, the failure test cases of Cart Pole (Fig.~\ref{fig:cartpole_reward}) are on the left side, those of Lunar Lander (Fig.~\ref{fig:lunarlander_reward}) appear in the lower-right region, \modify{and }{}those of Mountain Car (Fig.~\ref{fig:mountaincar_reward}) are concentrated near the center, close to $(-0.3,0)$\add{, and those of Humanoid (Fig.~\ref{fig:humanoid_reward}) are mainly distributed in the lower-right and the upper-left regions}. 

We observe that in Cart Pole (Fig.~\ref{fig:cartpole_reward}), the rewards of all passing test cases are identical, meaning the reward signal offers no guidance for distinguishing between passing and failing cases. In this setting, PRT and RT achieve the best performance. 
In Lunar Lander (Fig.~\ref{fig:lunarlander_reward}), although the rewards vary across the input domain, the passing test cases in the central region yield low rewards, which may mislead methods guided by reward signals. In this case, reward-driven baselines perform much better than those in Cart Pole, yet still fall short of PRT.
In Mountain Car (Fig.~\ref{fig:mountaincar_reward}), passing test cases exhibit diverse rewards. Fortunately, the failure region is surrounded by areas of low reward, allowing MDPFuzz and QD to efficiently detect failures. Nevertheless, PRT consistently ranks second in terms of the number cost among all methods.
\add{In Humanoid (Fig.~\ref{fig:humanoid_reward}), part of the failure test cases are located near the medium-reward test cases (green points), particularly in the upper-left region. However, these failure test cases are also surrounded by high-reward ones. As a result, MDPFuzz and CureFuzz can detect failures quickly in some cases, but in others it requires significantly more testing cost to detect failures.}

\begin{tcolorbox}
\textbf{Answer to RQ1:} \modify{PRT can effectively and efficiently detect failures with respect to both the number cost and the time cost.}{Compared with RT, PRT reduces the testing cost by at least 50\% across all testing subjects.} \modify{Its}{Compared with DRL-based testing techniques, PRT's} advantage is particularly evident in the environments where reward signals are either uninformative or misleading\modify{ for reward-guided algorithms}{}.
\end{tcolorbox}

\subsection{RQ2: Diversity of Test Cases}
\begin{figure}[htbp]
  \centering
  \subfloat[Cart Pole\label{fig:cartpole_seeds}]{
    \begin{minipage}[b]{1.0\textwidth}
      \centering
      \includegraphics[width=0.95\textwidth]{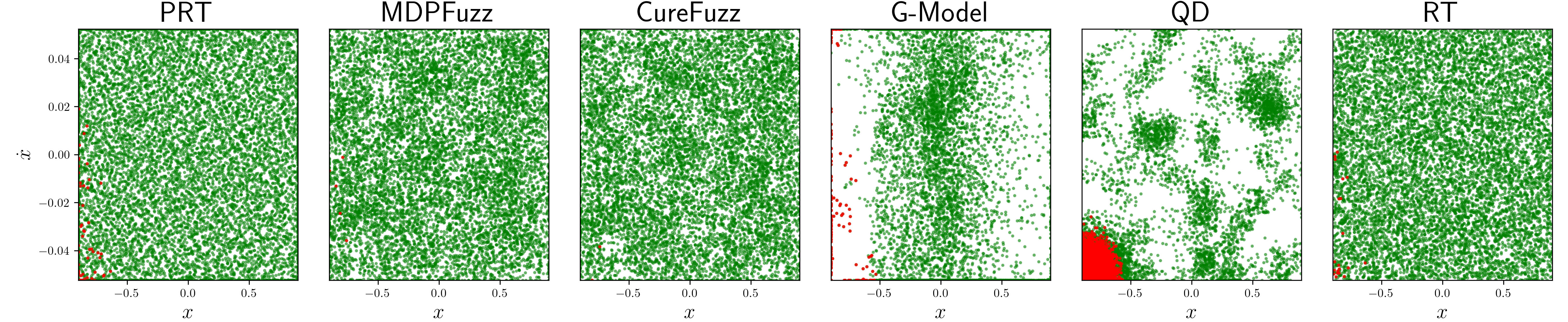} \\
      \includegraphics[width=0.95\textwidth]{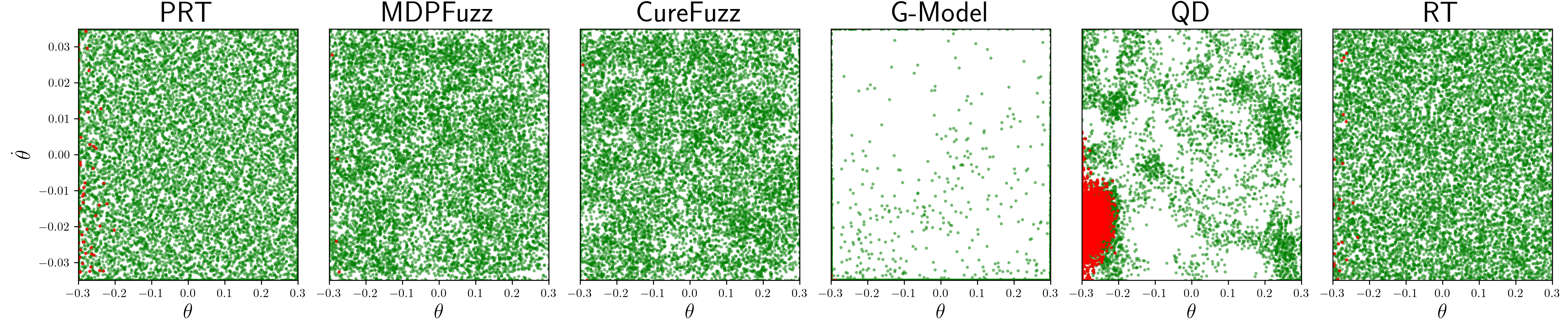}
    \end{minipage}
  }
  \hfill
  \subfloat[Lunar Lander\label{fig:lunarlander_seeds}]{
    \includegraphics[width=0.95\textwidth]{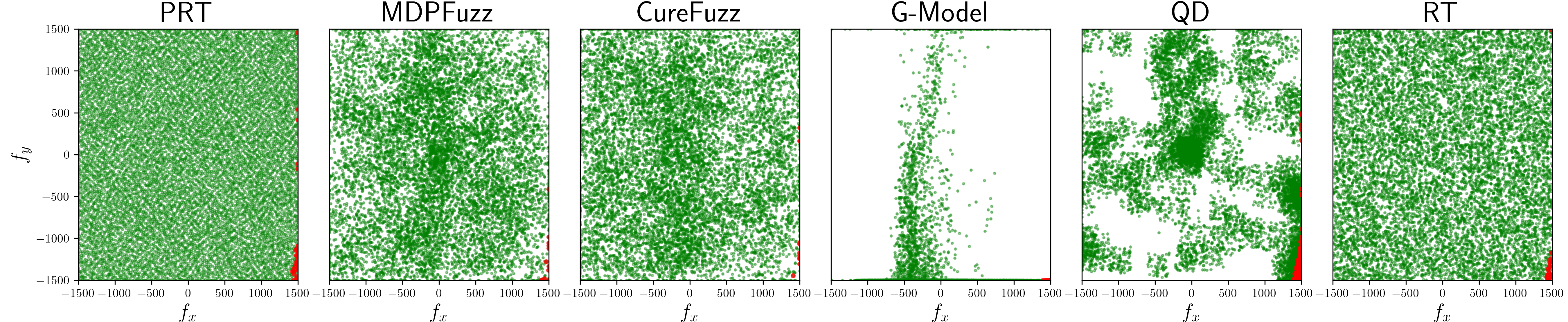}
  }
  \hfill
  \subfloat[Mountain Car\label{fig:mountaincar_seeds}]{
    \includegraphics[width=0.95\textwidth]{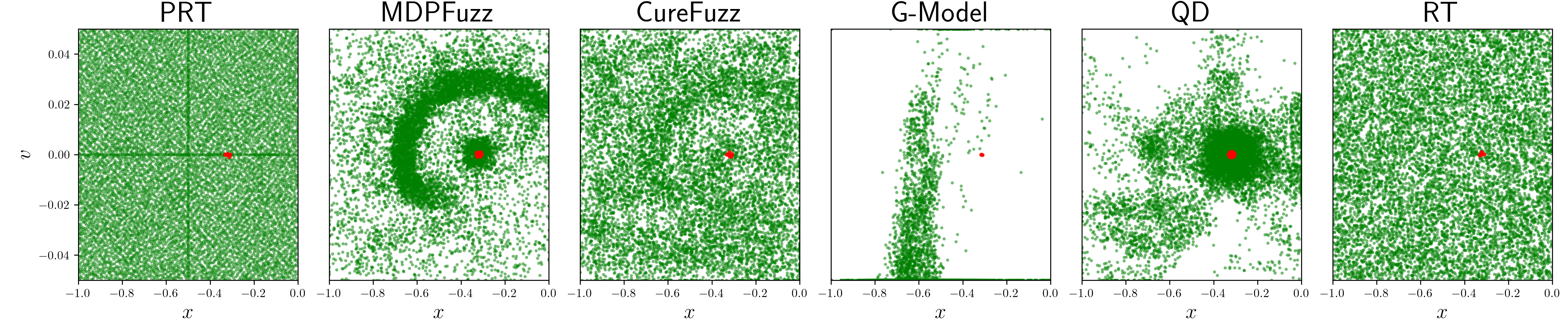}
  }
  \hfill
  \subfloat[Humanoid\label{fig:humanoid_seeds}]{
    \begin{minipage}[b]{1.0\textwidth}
      \centering
      \includegraphics[width=0.95\textwidth]{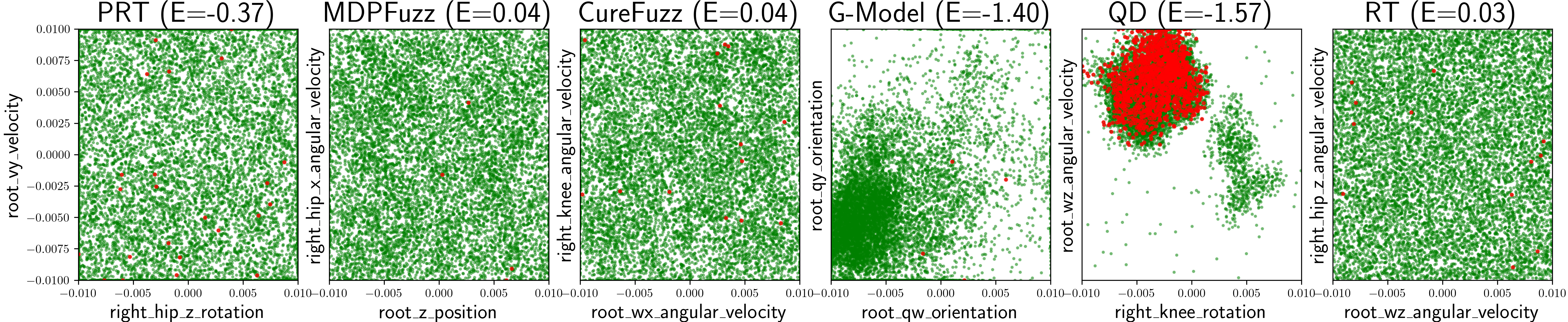} \\
      \includegraphics[width=0.95\textwidth]{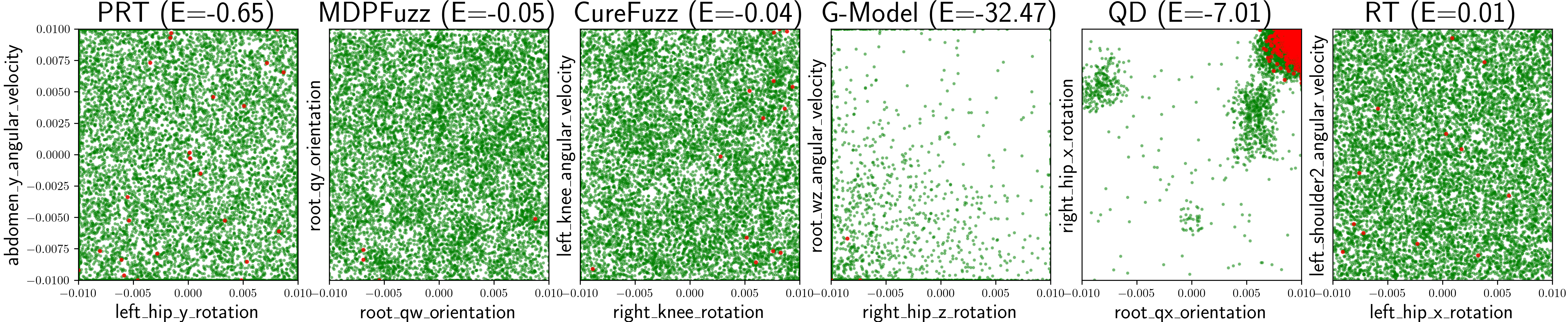}
    \end{minipage}
  }
  \vspace{-0.2cm}
  \caption{Test case distributions of different methods}
  \label{fig:seedspattern}
\end{figure}

\modify{}{We leverage entropy (Eq.~\ref{eq:entropy}) to evaluate the diversity of generated test cases by different methods. Each method is required to generate 10,000 test cases, and every experiment is repeated 20 times. To mitigate the threat of conclusion validity, we show the results with hyperparameters $k=5$, $k=10$ and $k=20$ in Eq.~\ref{eq:entropy}. We also report the standard variance of each experiment. 
Beyond such distance metric, we also visualize the test case distribution of different methods as Fig.~\ref{fig:seedspattern}. On one hand, this can implement the distance metric, making our conclusion more reliable. On the other hand, more importantly, that visualization intuitively shows the exploring strategies of different methods, enabling deeper analysis on their advantages and disadvantages. Note that, during the 20 experiments, we select the one with the lowest number cost finding the first failure for visualization, where the green and the red points represent the passing and the failure test cases respectively. }

\begin{table}[h]
\centering
\caption{Entropy of 10,000 generated test cases under different $k$}
\label{tab:entropy}
\begin{tabular}{lcrrrr}
\toprule
Method & $k$ & Cart Pole & Lunar Lander & Mountain Car & Humanoid \\
\midrule
\textbf{PRT} & \multirow{6}{*}{5} & $\bm{0.5266 \pm 0.0043}$ & $\bm{0.2306 \pm 0.0014}$ & $\bm{0.2120 \pm 0.0018}$ & $\bm{14.3975 \pm 0.4199}$ \\
MDPFuzz & & $-0.6569 \pm 0.0205$ & $-0.0446 \pm 0.0092$ & $-0.5768 \pm 0.0184$ & $-40.4884 \pm 0.6664$ \\
CureFuzz & & $-0.6567 \pm 0.0203$ & $-0.0289 \pm 0.0129$ & $-0.0320 \pm 0.0135$ & $-31.1420 \pm 0.6763$ \\
G-Model & & $-22.6375 \pm 2.6734$ & $-9.6081 \pm 0.7587$ & $-10.0524 \pm 1.0541$ & $-59.9535 \pm 2.8108$ \\
QD & & $-4.4818 \pm 0.4454$ & $-1.2159 \pm 0.1458$ & $-1.2140 \pm 0.1607$ & $-83.1634 \pm 0.7219$ \\
RT & & $0.1765 \pm 0.0068$ & $0.0157 \pm 0.0053$ & $0.0157 \pm 0.0053$ & $14.2386 \pm 0.0387$ \\
\midrule
\textbf{PRT} & \multirow{6}{*}{10} & $\bm{0.4513 \pm 0.0028}$ & $\bm{0.1054 \pm 0.0015}$ & $\bm{0.0880 \pm 0.0019}$ & $\bm{14.9516 \pm 0.3636}$ \\
MDPFuzz & & $-0.3097 \pm 0.0210$ & $-0.0360 \pm 0.0082$ & $-0.5727 \pm 0.0176$ & $2.3939 \pm 0.5797$ \\
CureFuzz & & $-0.3093 \pm 0.0203$ & $-0.0104 \pm 0.0082$ & $-0.0148 \pm 0.0102$ & $11.6711 \pm 0.3869$ \\
G-Model & & $-22.0423 \pm 2.7388$ & $-9.0635 \pm 0.7885$ & $-9.5217 \pm 1.0936$ & $-52.3097 \pm 2.8589$ \\
QD & & $-4.3464 \pm 0.4261$ & $-1.1395 \pm 0.1427$ & $-1.1994 \pm 0.1615$ & $-80.6062 \pm 1.2369$ \\
RT & & $0.2142 \pm 0.0071$ & $0.0215 \pm 0.0057$ & $0.0215 \pm 0.0057$ & $14.6885 \pm 0.0383$ \\
\midrule
\textbf{PRT} & \multirow{6}{*}{20} & $\bm{0.4044 \pm 0.0027}$ & $\bm{0.0911 \pm 0.0009}$ & $\bm{0.0731 \pm 0.0010}$ & $\bm{15.5194 \pm 0.3137}$ \\
MDPFuzz & & $-0.0061 \pm 0.0187$ & $-0.0292 \pm 0.0077$ & $-0.5719 \pm 0.0177$ & $14.6547 \pm 0.1209$ \\
CureFuzz & & $-0.0049 \pm 0.0177$ & $0.0012 \pm 0.0083$ & $-0.0039 \pm 0.0073$ & $14.8709 \pm 0.0826$ \\
G-Model & & $-21.2772 \pm 2.7295$ & $-8.5240 \pm 0.8100$ & $-9.0049 \pm 1.1442$ & $-42.7788 \pm 2.9618$ \\
QD & & $-4.1132 \pm 0.3829$ & $-1.0618 \pm 0.1529$ & $-1.1918 \pm 0.1622$ & $-77.3323 \pm 1.8267$ \\
RT & & $0.2589 \pm 0.0069$ & $0.0290 \pm 0.0037$ & $0.0290 \pm 0.0037$ & $15.1581 \pm 0.0371$ \\
\bottomrule
\end{tabular}
\end{table}

\modify{}{From Tab.~\ref{tab:entropy}, PRT consistently achieves the highest diversity across all environments, indicating that the test cases generated by PRT are more well-dispersed and uniformly distributed in the input space.
This observation is further supported by the visualizations in Fig.~\ref{fig:lunarlander_seeds} and Fig.~\ref{fig:mountaincar_seeds}, where PRT produces a more evenly distributed set of test cases than RT. In the other high-dimensional environments, though diversity of PRT is high in Tab.~\ref{tab:entropy}, Fig.~\ref{fig:seedspattern} cannot reveal the dimensional combinations, resulting in little difference on the two-dimensional prejections. 
Moreover, as shown in Fig.~\ref{fig:mountaincar_seeds}, PRT tends to focus more on the middle region of the input domain due to the mapping defined in Eq.~\ref{eq:mapping} in the Mountain Car environment. However, since the failure region in Mountain Car deviates from the center of the input space, this bias reduces the efficiency of PRT in detecting the first failure, reported in Tab.~\ref{tab:f1}.}

\add{For Humanoid, since the input domain has 47 dimensions, we enumerate all $\binom{47}{2} = 1081$ dimension pairs and select the two pairs with the highest and lowest entropy among \textbf{10,000 test cases}, as reported in Fig.~\ref{fig:humanoid_seeds}. From Fig.~\ref{fig:seedspattern}, we can observe that different algorithms exhibit distinct preferences.} MDPFuzz seems to be more interested in the middle region in Lunar Lander (Fig.~\ref{fig:lunarlander_seeds}) and the circle region in Mountain Car (Fig.~\ref{fig:mountaincar_seeds}), which fit the reward pattern shown in Fig.~\ref{fig:lunarlander_reward} and Fig.~\ref{fig:mountaincar_reward}. It corresponds to the best number cost in Tab.~\ref{tab:f1} Mountain Car environment. 
\modify{CureFuzz performs the highest diversity except for PRT and RT.}{CureFuzz achieves the third-highest diversity, following PRT and RT.} 
\modify{}{G-Model exhibits very low diversity. He et al. observed that G-Model tends to concentrate on boundary regions \cite{curefuzz}. Moreover, a GitHub \href{https://github.com/lizhuo-1994/mdp_testing/issues/1}{issue} reported that the generative model fails to converge during training, a problem we also encountered, leading to test cases outside the input domain. Following the solution proposed by Li et al., we project invalid test cases onto the boundary of the input domain.}
QD demonstrates a low diversity but finds many failures. That is because QD randomly samples a cluster of test cases but always selects the one with the lowest reward for executing. \add{Upon detecting a failure, QD concentrates on mutating the failure test case.}

Both G-Model and QD take an aggressive strategy---they focus on specific domains with a large number of test cases---but they may miss some failures. G-Model miss failures in Mountain Car. QD does not find the failures around $(x,\dot{x})=(-0.9,0.0)$ in Cart Pole environment (the middle left domain of the first line). Moreover, even if we add more test cases, it is still difficult for these methods to find the failures in such domain. That highlights the importance of test case diversity, which can help us miss fewer failures in testing. 

\begin{tcolorbox}
\textbf{Answer to RQ2:} PRT achieves the best diversity of test cases among all methods. \modify{More specifically, its coverage is at least 6\% higher than that of RT, the second-best method in our experiments.}{Notably, PRT consistently performs better than RT across all testing subjects.}
\end{tcolorbox}

\subsection{RQ3: Failure Patterns}
In this section, we examine the failure patterns exhibited by different agents. Specifically, we empirically study the regions of the input domain where failures occur. \add{Chen et al classified the patterns of failure test cases into three categories: \textit{point}, \textit{strip} and \textit{block} patterns \cite{art}. However, the dimensionality of Humanoid input domain is 47. Limited number of evenly distributed test cases cannot reveal the failures' geometric characteristics. Even if we only place two points in every dimension, the combinations can be $2^{47}\approx 10^{11}$. So we just explore failure patterns of the other three environments.} For Cart Pole (discrete version), we select agents trained with A2C \cite{a2c}, QRDQN \cite{qrdqn}, and PPO \cite{ppo}; for Lunar Lander (continuous version), SAC \cite{sac}, TD3 \cite{td3}, and PPO \cite{ppo}; and for Mountain Car (continuous version), DDPG \cite{ddpg}, TRPO \cite{trpo}, and TQC \cite{tqc}. In total, eight algorithms are considered to ensure the robustness of our conclusions. Testing is performed using PRT, which achieves the highest diversity according to Tab.~\ref{tab:entropy}. For each testing subject, PRT generates up to 100,000 test cases to reduce the likelihood of missing failures.

\add{Note that PRT is characterized by: (1) prioritizing specific regions (boundary in Cart Pole, Lunar Lander and Humanoid), and (2) uniform distribution. That implies PRT is significantly effective for failures aligned with the manually defined regions and \textit{block-shaped} failures.}

\begin{figure}[htbp]
  \centering
  \subfloat[Cart Pole\label{fig:cartpole_failure}]{
    \includegraphics[width=0.98\textwidth]{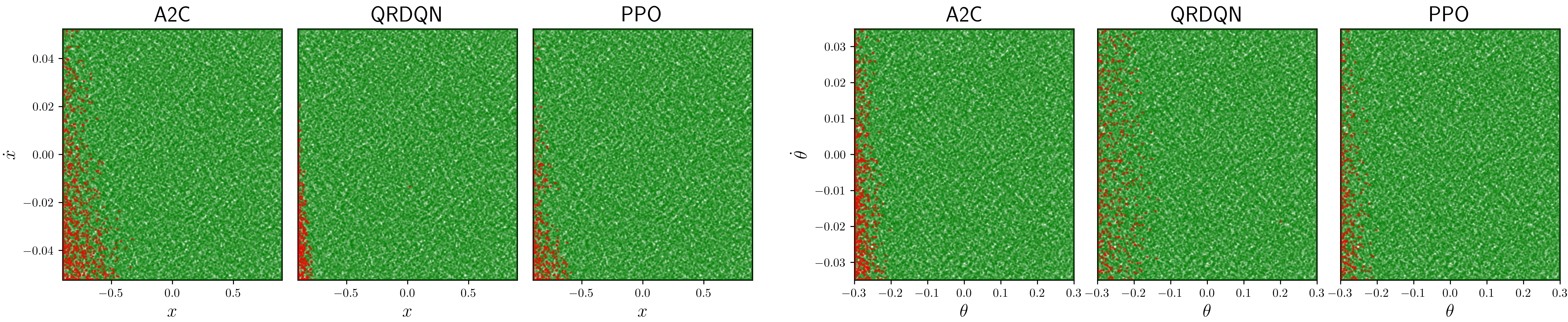}
  }
  \hfill
  \subfloat[Lunar Lander\label{fig:lunarlander_failure}]{
    \includegraphics[width=0.48\textwidth]{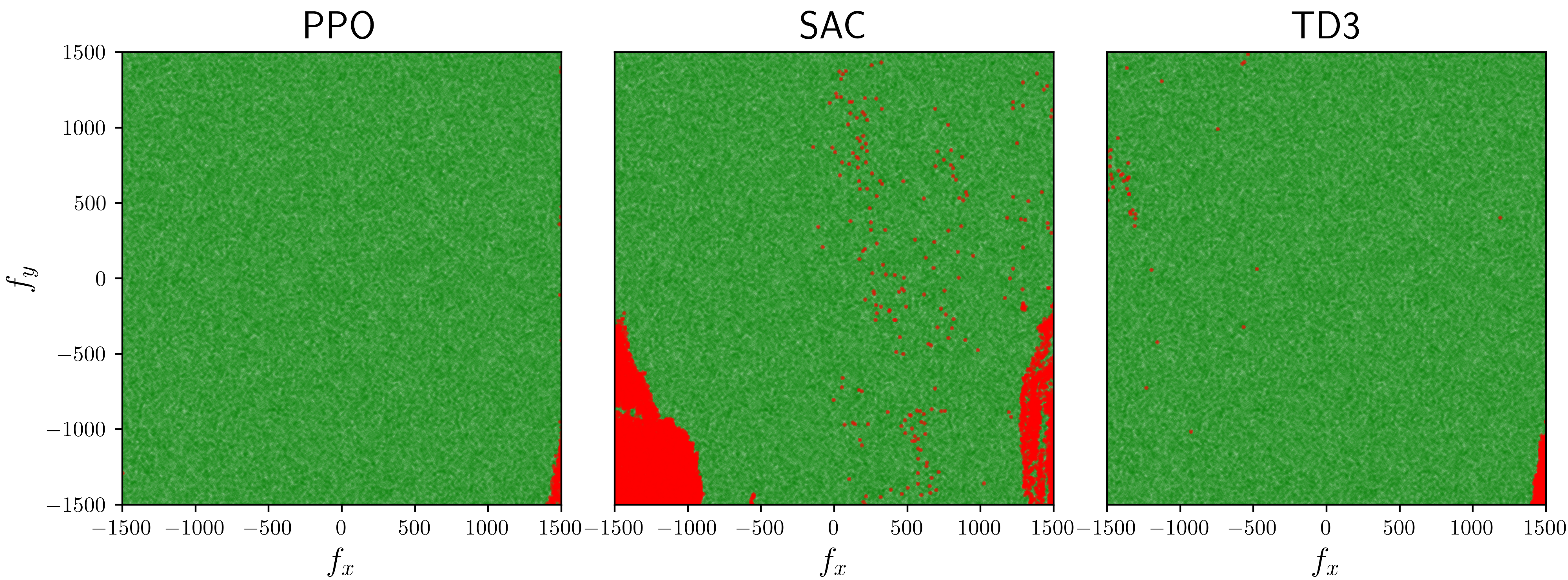}
  }
  \subfloat[Mountain Car\label{fig:mountaincar_failure}]{
    \includegraphics[width=0.48\textwidth]{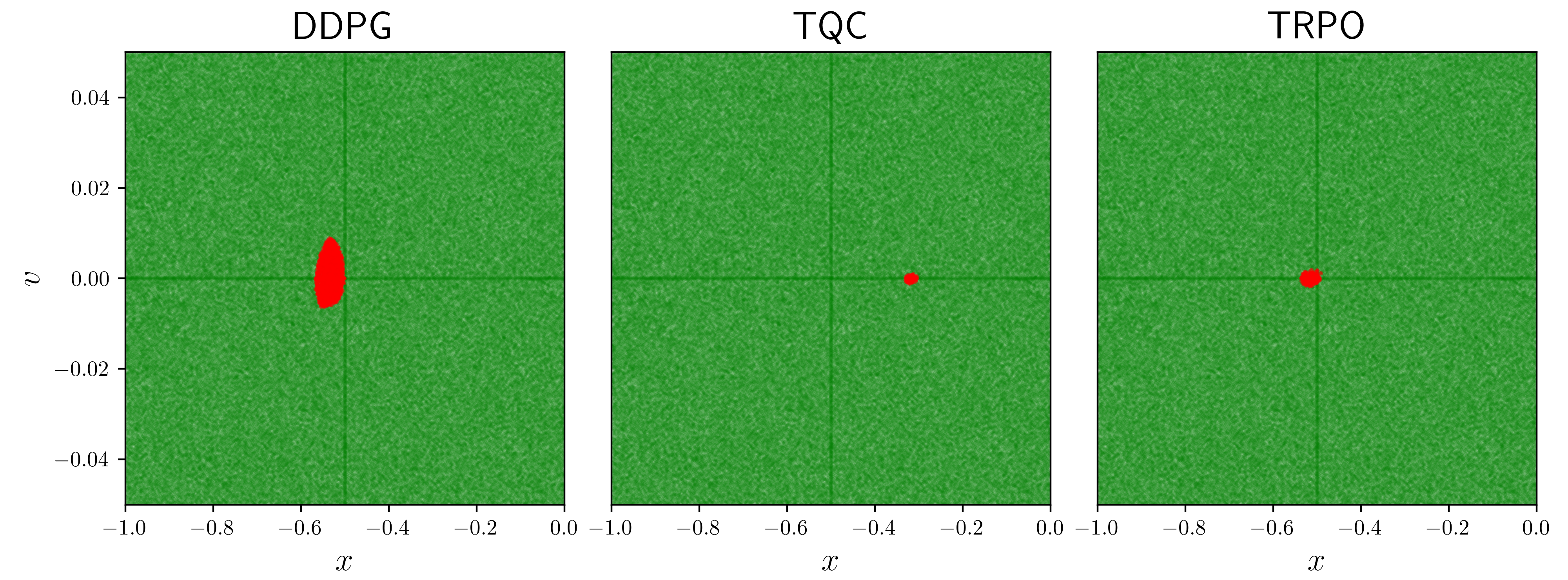}
  }
  \caption{Failure patterns of different agents.}
  \label{fig:failure_patterns}
\end{figure}

Fig.~\ref{fig:failure_patterns} illustrates the distribution of passing test cases (green) and failure test cases (red). \modify{We derive two main observations:

\begin{enumerate}[leftmargin=2em]
\item Most failures are aligned with the task-induced failure insights, forming \emph{block-shaped} clusters in specific regions.  
\item Failures and passing test cases may also appear sparsely, where isolated failures are surrounded by passes and isolated passes are surrounded by failures, forming \emph{point-shaped} patterns in local regions not algined with the task-induced failure insights. 
\end{enumerate}}{}
\modify{Though the block-shaped pattern can be exploited to efficiently uncover a large number of failures, point-shaped failures pose additional challenges, as they are difficult to expose and can be overlooked even in dense regions of passing test cases. Thus, the existence of point-shaped failures implies that no amount of local passing cases guarantees the absence of failures in the gaps between them. 

On one hand, block-shaped failures align well with task-induced failure insights, which PRT leverages to detect failures efficiently. On the other hand, the point-shaped failure pattern underscores the importance of uniform distribution, which is inherently provided by PRT. These characteristics together enable PRT to detect failures in DRL agents both effectively and efficiently.}{We find that the testing subjects exhibit \emph{block-shaped} (clustered) failures aligned with the task-induced failure insights. That is the reason why PRT achieve significant effectiveness in Tab.~\ref{tab:f1}. Moreover, we observe that such aligned and \emph{block-shaped} patterns appear across different training algorithms in the same environment, which implies the task-induced failure insights are largely environment-dependent and agent-independent and such failure insight is a general tool for testing DRL agents. However, there are also \emph{point-shaped} (dispersed) failures not aligned with the failure insights. Though not efficiently, PRT detects them by uniform distribution.}

\begin{tcolorbox}
\textbf{Answer to RQ3:} \modify{We observe two main failure patterns in DRL agents: block-shaped clusters and point-shaped sparsity. The block-shaped failures align with task-induced failure insights, allowing PRT to detect them efficiently. Meanwhile, thanks to PRT's broad coverage, point-shaped failures can also be effectively exposed.}{We empirically observe that there are \emph{block-shaped} failures aligned with the task-induced failure insights across different training algorithms in the same environment, where PRT detects such failures by testing priority. There are also \emph{point-shaped} failures not aligned with the task-induced failure insights. PRT can detect them by uniform distribution.}
\end{tcolorbox}

\section{Discussion}
\textbf{The applicable conditions of PRT.} Though PRT is capable of generating evenly distributed points within a hypercube (defined as Eq.~\ref{eq:hypercube}), certain techniques can be applied to broaden its applicability.

\begin{equation}
  \mathcal{D} = \{x\in\mathbb{R}^m | x^i\in[inf_i,sup_i], i=1,2,\cdots,m\}
  \label{eq:hypercube}
\end{equation}

A hypercube is essentially a domain in which each dimension is independently bounded. To satisfy the boundedness assumption, a sufficiently large value can be chosen to approximate the infimum or the supermum. For the independence assumption, if the domain has a complex shape, a larger hypercube can be used to enclose the domain. Points generated outside the target domain are filtered out, while those within are retained. 

\noindent\textbf{How to understand the hyperparameters $\tau$ and $\lambda$?} \modify{PRT assumes the failure domain is a hypercube with side length $\tau$ and it tries to avoid the case that two test cases are within a $\tau$-length hypercube.}{PRT assumes that the failure region is a hypercube with side length $\tau$, and it seeks to prevent two test cases from residing within the same $\tau$-length hypercube. Nevertheless, owing to randomness, it is challenging to strictly determine whether two points lie within such a region. Hence, assuming the existing points are evenly distributed, we employ a ratio model as an approximation.} As PRT prioritizes testing a specific domain, $\lambda$ exhibits confidence in that task-induced failure insight. If $\lambda=1$, the generating process follows a uniform distribution and that means \modify{the expert knowledge is unreliable.}{the task-induced failure insight plays little role in the test case generation process.}

\noindent\textbf{Can PRT be applied to test other software systems?} The answer is yes. PRT is a failure-based testing method. We use PRT to test DRL agents because we can easily locate their failure-prone input domains. If we know about the size or the potential failure domain of other softwares, we can also leverage PRT to test them. 

\noindent\textbf{Why some methods perform worse than RT in Tab.~\ref{tab:f1}?} Firstly, the evaluation metrics differ: while most methods are assessed by the total number of failures detected, we measure the cost for finding the first failure. Secondly, the methods have different search preferences. Many are effective at uncovering clustered failures, whereas RT distributes its effort across the entire input domain, avoiding excessive testing in any local region. At last, we have explained that in some environments they lack effective reward guidance, which will make their performance worse.

\section{Threats to Validity}
\subsubsection*{Internal Validity}  
A potential threat to internal validity lies in the setting of hyperparameters. For PRT, we describe the hyperparameters in Sec.~\ref{hyperparameter}. For the baselines, we strictly follow the settings reported by their authors whenever available. For MDPFuzz \cite{mdpfuzz}, Mazouni et al. confirmed that the hyperparameters are set as $K=10,\tau=\gamma=0.01$ \cite{2024policy}. For CureFuzz \cite{curefuzz}, the authors did not report the hyperparameter settings. We therefore examined its \href{https://github.com/soarsmu/CureFuzz}{GitHub repository} and found that the settings differ from the formulas in the paper. In our experiments, we followed the GitHub settings, normalized the rewards of different environments to $[0,1]$, and set $\alpha=5,\beta=0.01,\gamma=1$ for all experiments. For G-Model \cite{gmodel}, all hyperparameters are clearly stated in the paper, and we adopt them directly. For QD \cite{qd}, we refer to its \href{https://github.com/QuentinMaz/QD_Based_Testing_RL}{GitHub repository}. As it also uses the Lunar Lander environment, we take the same settings for that task. For \modify{Cart Pole and Mountain Car}{Cart Pole, Mountain Car and Humanoid}, we derive the settings by analogy from those of Taxi.  

\subsubsection*{External Validity}  
A limited number of testing subjects poses a threat to external validity. One concern is how to design task-induced failure insights for other environments. As discussed in Sec.~\ref{expertknowledge}, such insights can be derived from the tasks\modify{ and reward designs}{}, which are intrinsic characteristics of DRL agents. Another concern is the generalizability of our findings. To mitigate this threat, we analyze the underlying mechanisms of different testing methods from multiple perspectives (e.g., reward patterns, test case distributions, and failure patterns), with the aim of explaining why the observed behaviors are expected, in order to generalize beyond the studied benchmarks.

\add{\subsubsection*{Construct Validity}
A threat to construct validity lies in the diversity metric. Eq.~\ref{eq:entropy} can be unstable due to large dimension $d$ or different hyperparameter $k$. However, $d$ only influcences the constant item, and that is equivalent when we only compare the entropy under the same environment and the same test case number. To mitigate the threat from $k$, we report $k=5,10,20$ in RQ2.
}

\subsubsection*{Conclusion Validity}  
A threat to conclusion validity may come from the randomness of test execution. To mitigate this, we repeat each experiment 20 times and report the averaged results. \add{In addition, we report ($p$-value, effect size) in RQ1, and standard variance in RQ2.} To further improve reproducibility, we fix the random seed in our open-source implementation.  


\section{Related Works}

\textbf{Fuzz testing.} Pang et al. proposed MDPFuzz \cite{mdpfuzz}, which leverages rewards to calculate seed sampling probabilities and employs a Gaussian Mixture Model (GMM) to estimate state freshness. Mazouni et al. verified that while the GMM strategy is effective, it is also time-consuming \cite{2024policy}. He et al. developed CureFuzz \cite{curefuzz}, a curiosity-driven fuzzing approach. Wan et al. introduced DRLFuzz \cite{drlfuzz}, a white-box fuzzing method that uses a coverage-guided algorithm and the Q-network gradient to mutate seeds for testing DRL agents.

\textbf{Search-based Testing.} Zolfagharian et al. proposed a white-box search-based testing method STARLA \cite{starla}. It leverages the Q-value of agents to encode the state sequence into a binary sequence and applies a random forest algorithm to predict the failure probability of that sequence as the fitness function. Biagiola and Tonella introduced Indago \cite{surrogate}, searching for failure environment configurations with a failure predictor to give the failure probability. However, both STARLA and Indago rely on training data---test cases and their corresponding passing or failure labels. \modify{However, off-learning DRL techniques, such as DQN \cite{dqn}, TD3 \cite{td3}, DDPG \cite{ddpg} and TQC \cite{tqc}, only need a data set and do not execute the training agents. The labels of test cases cannot be obtained in the training phase, so STARLA and Indago cannot test off-learning DRL agents. }{}Mazouni et al. characterize a sequence with a 2-dimensional array as the agents' behavior, so as to guide genetic search for failures \cite{qd}. Ma et al. proposed MASTest \cite{mastest}, a diversity-guided testing method for multi-agent systems. 

\textbf{Generative-based Testing.} Li et al. introduced the diffusion model to generate test cases \cite{gmodel}. In every epoch, random test cases are sampled to fine-tune the diffusion model, which then generates a batch of new test cases for evaluation.

\section{Conclusion and Future Work}

In this work, we focus on well-trained DRL agents and propose a failure-based testing technique PRT. Our key point is that DRL agents are task-oriented and we can intuitively have the failure insights about which task is harder and the corresponding region has a higher failure probability. With the task-induced failure insights, we leverage PRT to prioritize testing the specific regions of the input domain, in order to uncover failures as fast as possible. Our empirical experiments show that PRT can effectively and efficiently find failures in terms of the number of test cases and the test execution time. Its advantage is particularly evident in the environments where reward signals are either uninformative or misleading for reward-guided algorithms. Furthermore, we investigate the failure patterns of several agents across different environments and we find that the task-induced failure insights are empirically aligned with the block-shaped failures while the characteristic of uniform distribution can help PRT to find the point-shaped failures.

In the future, we plan to apply PRT to testing other DRL agents \add{according to their specific failure insights }as well as other types of software. \add{For example, the failure-prone regions of codes can be the boundary or branch conditions in the program, which can be served as the failure-insights for PRT to test. }Besides, since PRT is efficient in finding the first failure and some methods can exploit specific failure characteristics, we will further investigate the feasibility of combining PRT with these methods, aiming to improve the effectiveness of detecting clustered failures while maintaining high efficiency and uniform distribution.

\section*{Acknowledgements}

This work was partially supported by the National Key R\&D Program of China (Grant No. 2024YFB33
11503) and the National Natural Science Foundation of China (Grant No. 62372021). We sincerely thank Tsong Yueh Chen (Swinburne University of Technology), Yuechen Li (Beihang University), Yi Cai (Beihang University), and Tianjie Zhou (Beijing Normal University) for helpful discussions and insightful comments.

\section*{Data Availability}

We open our source code at \href{https://github.com/Avagnes/PRT-DRL-Experiments/tree/main}{https://github.com/Avagnes/PRT-DRL-Experiments/tree/main} to show more details and help for further researches. 

\bibliographystyle{ACM-Reference-Format}
\bibliography{ref}

\end{document}